\documentclass[aps,prb,twocolumn,floats,superscriptaddress,epsfig]{revtex4-2}
\usepackage[T1]{fontenc}
\usepackage{amssymb}
\usepackage{amsbsy}
\usepackage{amsmath}
\usepackage{epsfig}
\usepackage{color}
\usepackage{float}
\usepackage[colorlinks,linktocpage,bookmarks=false,citecolor=blue,linkcolor=red,urlcolor=blue]{hyperref}

\newcommand{\ing}{\includegraphics}

\newcommand{\beq}{\begin{equation}}
\newcommand{\eeq}{\end{equation}}
\newcommand{\bea}{\begin{eqnarray}}
\newcommand{\eea}{\end{eqnarray}}

\newcommand{\ket}[1]{|#1\rangle}

\newcommand{\ave}[1]{\langle #1 \rangle}

\graphicspath{{./figures/}}

\begin{document}

\title{Floquet engineering of binding in doped and photo-doped Mott insulators}
\author{Madhumita Sarkar}
\affiliation{Jo{\v z}ef Stefan Institute, SI-1000, Ljubljana, Slovenia}

\author{Zala Lenar{\v c}i{\v c}}

\affiliation{Jo{\v z}ef Stefan Institute, SI-1000, Ljubljana, Slovenia}

\author{Denis Gole{\v z}}
\affiliation{Jo{\v z}ef Stefan Institute, SI-1000, Ljubljana, Slovenia}
\affiliation{Faculty of Mathematics and Physics, University of Ljubljana, 1000
Ljubljana, Slovenia}

%\date{\today}

\begin{abstract}

We investigate the emergence of bound states in chemically and photo-doped Mott insulators, mediated by spin and $\eta$-pairing fluctuations within both 2-leg ladder and 2D systems. To effectively describe the photo and chemically doped state on the same footings, we employ the Schrieffer-Wolff transformation, resulting in a generalized $t$-$J$ model. Our results demonstrate that the binding energies and localization length in the chemically and photo-doped regimes are comparable, with $\eta$-pairing fluctuations not playing a crucial role. Furthermore, we show that manipulating the binding is possible through external periodic driving, a technique known as Floquet engineering, leading to significantly enhanced binding energies. We also roughly estimate the lifetime of photo-doped states under periodic driving conditions based on the Fermi golden rule. Lastly, we propose experimental protocols for realizing Hubbard excitons in cold-atom experiments.
\end{abstract}

\maketitle

\section{Introduction}
\label{intro}

Doped Mott insulators are one of the most intensively studied solid-state systems due to their significance for the high-temperature superconductors~\cite{lee2006} and a plead of intriguing exotic phases including pseudogap~\cite{warren1989,Norman2005,wu2018},stripes~\cite{Tranquada1995,corboz2014,Wietek2021}, etc. One of the basic questions is how chemically doped charge carriers interact with the antiferromagnetic background, and it was proposed early on that carriers can form bound pairs whose binding energy originated from the shared distortion of the spin background~(string-based pairing)~\cite{dagotto1994,maier2006,maier2008,Jaklic2000,trugman1988,bonca1989,Chernyshev1998,bonca2007}. For the binding of charge carriers, the dimensionality of the system plays a particularly important role as in the purely 1D systems spin-charge separation prevents pairing~\cite{ogata1990,shiba1991,giamarchi2003quantum,essler2005one}, making ladder~\cite{dagotto1994,dagotto1992} and 2D systems~\cite{Jaklic2000,bonca2007} minimal setups for studying of bound pairs. Experimental progress in cold atoms simulators enabled a direct observation of antiferromagnetic correlations~\cite{mazurenko2017},  dressed spin polarons~\cite{Koepsell2019} and pairing strings~\cite{chiu2019,Bohrdt2019}. Furthermore, combining cold atom experiments and strong external fields enables a unique opportunity to manipulate microscopic parameters, like hopping and superexchange amplitude~\cite{Trotzky2008,dimitrova2020}. Recent examples include enhanced binding of holes under a strong DC field in regimes where superexchange becomes comparable to the hopping integrals~\cite{Hirthe2023,bohrdt2022}, a situation not available in solid-state setups. These ideas stimulate exploring whether one could employ periodic~(Floquet) driving~\cite{Bukov2015,Bukov2016,Mentink2015}, whose ability to manipulate microscopic parameters was already proved in cold-atoms experiments~\cite{gorg2018}, to stabilize bound pairs in doped Mott insulators even further.

Photo doping of Mott insulators is an emerging direction to form exotic phases of matter~\cite{murakami2023rev,Sentef2021}. While the photo-doped systems are metastable, their lifetime can be exponentially long-lived for large gap Mott insulators~\cite{Sensarma2010a,Strohmaier2010,kollath2007,Zala2013PRL,lenarcic13}. These states can exhibit novel non-thermal correlations, like the formation of Hubbard excitons~\cite{tohyama2006,jeckelmann2003,Zala2013PRL,lenarcic13,Bittner2020,Sugimoto2023,matsueda2004,Okamoto2019SciAdv,Mehio2023} or $\eta$-pairing~\cite{rosch2008,kaneko2019,murakami2022,ueda2023,kaneko2020,ejima2020,tindall2019,shirakawa2020}, etc. In the high-dimensional systems, it was shown that $\eta$-pairing fluctuations can lead to the condensation and form a superfluid~\cite{Li2020}, while in the one-dimensional setup a spin-charge-$\eta$ separation takes place~\cite{murakami2023b}. An important open question is to establish what are the properties of photo-doped states beyond the two extreme case studies, like the ladders and 2D situation, as these are the most relevant experimental situations.

In this work, we compare the formation of a bound pair for chemically and photo-doped systems in the ladder and 2D geometries. To describe the chemically doped and metastable photo-doped systems on the same footing, we resort to the canonical transformation, which perturbatively decoupled sectors with different numbers of holons and doublons.  We establish that the pairing due to spin and  $\eta$ fluctuations does not compete or actively cooperate but rather leads to similar bindings between charge carriers in chemically and photo-doped systems. We provide an approximate measure of the binding energy for both chemically and photo-doped situations.  Further, we show that periodic driving with an electromagnetic field enables an efficient manipulation of binding energies with its substantial increase for ladder and 2D systems and more localized charge distribution in real space. For photo-doped systems, we estimate the upper bound for the doublon-holon recombination rate in the presence of Floquet driving and proposed driving regimes where the exciton binding energies are large and lifetimes sufficiently long. Finally, we comment on preparing photo-doped states in cold atom experiments by either chirped driving across the gap or adiabatic deformation of the lattice potential from band to Mott insulator.

\section{Model and method}
\label{hamil}
We consider the Fermi Hubbard model
\begin{eqnarray}\label{eq::Hubbard}
 H &=& -t \sum_{ \ave{i,j} , \sigma} ( c^{\dagger}_{i \sigma} c_{j \sigma} + c^{\dagger}_{j \sigma } c_{i \sigma } ) + U \sum_{j} n_{ j \uparrow } n_{ j \downarrow},   
\end{eqnarray}
where sum $\ave{i,j}$ runs over nearest neighbour pairs of sites, $\sigma$ is the spin index denoting either up or down spin, $t$ is the hopping amplitude and $U$ the interaction strength.
We will focus on the strong interaction limit $U\gg t$, in which case the ground state at half-filling is the Mott insulating state. We will consider two different geometries with $N$ sites: (i) quasi one-dimension two-leg ladder with periodic boundary conditions along the elongated (chain) direction and (ii) two-dimensional lattice with periodic boundary conditions along both directions. The aim of our study is to understand and compare charge pairing and the role of spin and $\eta$ fluctuations in photo and chemically doped correlated systems.

While chemically doped systems are  stable, the photo-doped situation for the large gap Mott insulator is only metastable with exponentially long lifetime~\cite{Strohmaier2010,Sensarma2010a,Zala2013PRL,Eckstein2011thermalization}. To treat the two scenarios on the same footing, we will perform a canonical Schrieffer-Wolf transformation which perturbatively decouples sectors with different number of effectively charged holons and doublons and obtain a generalized version of celebrated $t$-$J$ Hamiltonian~\cite{MacDonald1988,Li2020,murakami2022}. In the case of the photo-doped system, by neglecting higher-order recombination terms, we approximate the metastable state with an equilibrium state. Conveniently, such a description reduces the computational complexity. We then estimate the Hubbard exciton lifetime by Fermi golden rule in analogy with Ref.~\onlinecite{Zala2013PRL}.

\begin{figure}[t!]
\includegraphics[width=0.8\columnwidth]{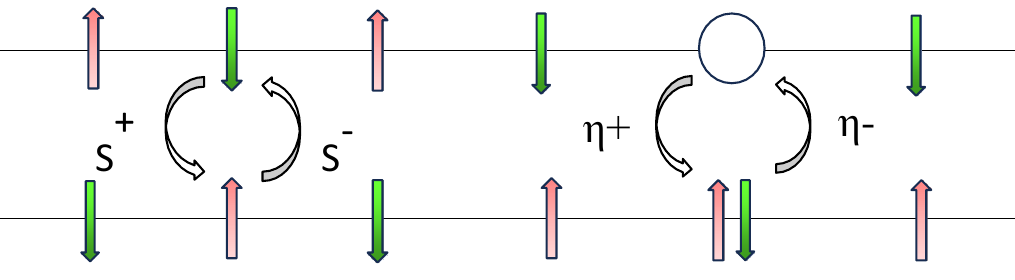}
\includegraphics[width=0.8\columnwidth]{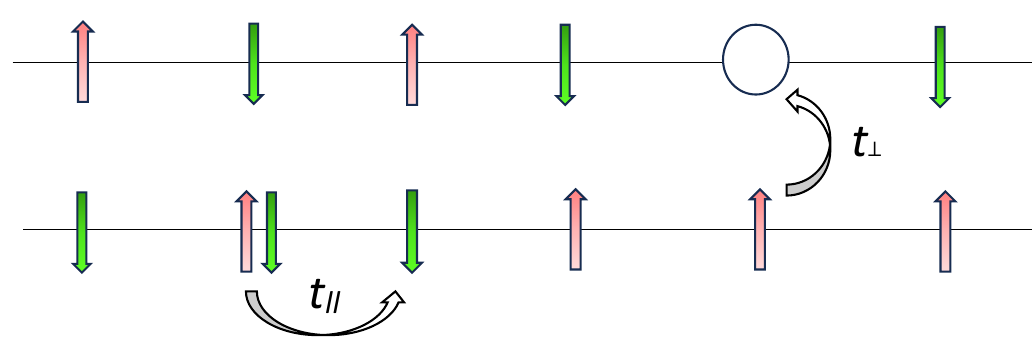}
\caption{Schematic representation of terms entering the effective model $H_{\text{eff}}$, see Eq.~\eqref{eq::Heff}. Top: the spin exchanges process, which exchanges the position of up and down spin; the $\eta$-exchange process, which exchanges the position of a doublon and a holon. Bottom: The hopping process of holon/doublon without changing the number of DH pairs.}
\label{fig:fig1}
\end{figure}

We denote the onsite interaction term in Eq.~\eqref{eq::Hubbard} with $H_0=U\sum_j n_{j\uparrow} n_{j\downarrow}$, and split the kinetic part into terms that preserves the number of doublon-holon (DH) pairs $T_0$,  that increases the number of DH pairs by one $T_1$, and that decreases the number of DH pairs by one $T_{-1}$. Following Ref.~\onlinecite{MacDonald1988}, the generator of the unitary transformation $S$ is chosen such that it transforms out the processes that change the number of DH pairs on the order $O(t)$ and retains them on the higher $O(t^2/U)$ level of expansion. %\textit{after truncation at the lowest order of $t^2/U$ expansion}. 
Hence, we choose $S=(T_1 - T_{-1})/U$ such that $T_1$ and $T_{-1}$ are cancelled (in Eq. ~\eqref{sw1}),
\begin{eqnarray}
\label{sw1}
H^{\prime}&=& e^{iS} H e^{-iS} \\
&=& H_0 + T_0 + T_1 + T_{-1} + [ i S, H] + \dots \nonumber\\
&=& H_0 + T_0 + \frac{1}{U} \big( [ T_1 , T_{-1} ] + [T_0 , T_{-1}] + [ T_1 , T_0 ] \big) + \dots \nonumber
\end{eqnarray}
In the following, we distinguish the hopping amplitudes in x- and y-direction and denote them as $t_{\parallel}$ and $t_{\perp}$ respectively.  For 2D model, $t=t_{\parallel}=t_{\perp}$, as can be easily imposed in the following expressions. In the limit $U\gg t_{\parallel},t_{\perp}$, we retain terms of the order $O(t_{\parallel}^2/U)$, $O(t_{\perp}^2/U$)  and neglect all higher orders, such that the effective Hamiltonian preserves the number of DH pairs
\begin{eqnarray}\label{eq::Heff}
H_{\text{eff}}=& \sum\limits_{\alpha\in\{\parallel,\perp\}} \left[ H_{\eta}^{\alpha}+H_{S}^{\alpha}+H_{H,hop}^{\alpha}+H_{D,hop}^{\alpha}\right]+H_0\nonumber.
\end{eqnarray}
The $\parallel$ terms act on the charge and spins along the x-direction (along the chain) whereas the $\perp$ terms act in the y-direction (along the rung for the two-leg model). When written out explicitly, the effective Hamiltonian has the following terms\cite{murakami2022}
\begin{eqnarray}
H_{S}^{\parallel(\perp)}&=&  J_{\parallel(\perp),S} \sum_{\langle i, j\rangle } \left(\boldsymbol{S}_{i } \cdot  \boldsymbol{S}_{j} -\frac{1}{4} \delta_{1,n_{i }  n_{j }}\right) ,\nonumber \\  
H_{\eta}^{\parallel(\perp)}&=&  -J_{\parallel(\perp),\eta} \sum_{\langle i,j\rangle}
\left(\boldsymbol{\eta}_{i} \cdot \boldsymbol{\eta}_{j } - \frac{1}{4}(1-\delta_{1,n_{i }})(1-\delta_{1,n_{j}})\right), \label{eq:H_eta}\nonumber\\
H_{D,hop}^{\parallel(\perp)}&=& -t_{\parallel(\perp)} \sum_{\langle i,j\rangle , \sigma} n_{i \bar{\sigma} } ( c^{\dagger}_{i \sigma } c_{j \sigma } + h.c. ) n_{j \bar{\sigma} },\label{eq:Dhop}\nonumber\\ 
H_{H,hop}^{\parallel(\perp)}&=& -t_{\parallel(\perp)} \sum_{\langle i,j\rangle , \sigma} \bar{n}_{i \bar{\sigma} } ( c^{\dagger}_{i \sigma } c_{j \sigma } + h.c. ) \bar{n}_{j \bar{\sigma} }, \label{eq:Hhop} \\ \nonumber
% H_{den} &=& \frac{1}{4}  \sum_{\langlei,j>, \sigma} ((J_{\parallel(\perp),\eta}(1-\delta_{1,n_{i }})(1-\delta_{1,n_{j}}))  \\ &-&  J_{\parallel(\perp),S} \delta_{1,n_{i }  n_{j }})\\ \nonumber 
 \label{htj}
\end{eqnarray}
where $J_{\parallel,S} = J_{\parallel,\eta} = \frac{4t_\parallel^2}{U}$ and $J_{\perp,S} = J_{\perp,\eta} = \frac{4t_\perp^2}{U}$
are the spin/$\eta$ exchange coupling parameters along the chain and rung, respectively. When of the same strength, we drop the subscript $S$ and $\eta$. The electron~(hole) density per spin is given by 
$n_{i,\sigma}~(\bar{n}_{i,\sigma}=1-n_{i,\sigma})$, and $\bar{\sigma}$ is the complement of $\sigma$, i.e., if $\sigma$ denotes up spin then $\bar{\sigma}$ denotes down spin and vice versa. The $\eta$ operators in terms of the fermion creation and annihilation operators can be expressed as $\eta_i^{+}=\eta_i^x+ i \eta_i^y= e^{i\boldsymbol{r}_i\cdot\boldsymbol{\pi}} c_{i,\uparrow}^{\dagger}c_{i,\downarrow}^{\dagger}$, where $\boldsymbol{r}_i$ is the vector to the $i$th site and $\boldsymbol{\pi}=\{\pi, \pi, ...\}$ is a vector of the dimension of the system with entries $\pi$. The $z$ component of the pseudospin is $\eta_i^z=\frac{1}{2}(1-n_i)$, $\eta_i^- = (\eta_i^{+})^{\dagger}$,  and the spin operators are defined as, $ \boldsymbol{S}_i= \frac{1}{2} c_{i \alpha}^{\dagger} \boldsymbol{\sigma}_{\alpha \beta} c_{i \beta}$, where $\boldsymbol{\sigma} = \{\sigma^x,\sigma^y,\sigma^z\}$ are  the standard Pauli matrices.

Different terms in the Hamiltonian have the following physical meanings:  $H_{S}^{\parallel (\perp)}$ is the spin interaction term, familiar from the standard t-J model; $H_{\eta}^{\parallel (\perp)}$ is the $\eta$-exchange term, which exchanges the position of a doublon and holon and causes the interaction between DH pair if they are on nearest neighbor sites; $H_{H,hop}^{\parallel (\perp)}$ and $H_{D,hop}^{\parallel (\perp)}$ represent hopping of holon and doublon conserving their number, respectively. These processes are illustrated in Fig.~\ref{fig:fig1}. In the sectors with no doublon quasi-particle, $H_{\text{eff}}$ reduces to the standard t-J model. $H_0$ is the on-site Coulomb interaction and leads to unimportant energy shift between the chemical and photo-doped system. 
Both chemically and photo-doped systems reduce to the solution of the ground state problem in different sectors and we employ the exact diagonalization using the Lanczos technique~\cite{Prelovsek_springer,Lanczos1950}. 
%For quasi one-dimension two-leg ladder, we consider periodic boundary conditions along the elongated (chain) direction. For two-dimensional lattice, we consider periodic boundary conditions along both directions. 
%For systems sizes considered, we typically use \ZL{?? - 140} Lanczos steps, which are sufficient to properly capture the ground states and their energy, while not yet causing the emergence of ghost states.

We should note that in the full expansion there appear also three site terms which are of two types: a) the holon/doublon correlated hopping terms of order $J_{\parallel(\perp)}$ which conserve the number of holon and doublon pairs and the analysis in Ref.~\onlinecite{murakami2022} showed that their effect is small in the dilute limit considered here, b) recombination terms of order $J_{\parallel(\perp)}$ which we will for a moment suppress to mimic the metastable state by an equilibrium problem. Later on, we will estimate the  lifetime of the metastable phase~(recombination time) by perturbative treatment of these 3-site recombination terms. 

\section{Results for two-leg ladder}\label{sec::results}
Our aim is to study the properties of bound doublon-holon  (Hubbard excitons) or two holon states and compare the binding of charged carriers in photo and chemically doped situation. An important goal is to understand the interplay of spin and $\eta$ fluctuations on the doublon-holon binding, in systems beyond one dimension, which is special due to the spin-charge and spin-$\eta$-charge separation~\cite{murakami2022,murakami2023b}.
 
We start our analysis with the simplest system which escapes the separation, namely the two-leg ladder system, and later extend the study to the two-dimensional situation.

\subsection{Exciton binding energy}
\label{pair}

\begin{figure}[t!]
\includegraphics[width=\columnwidth]{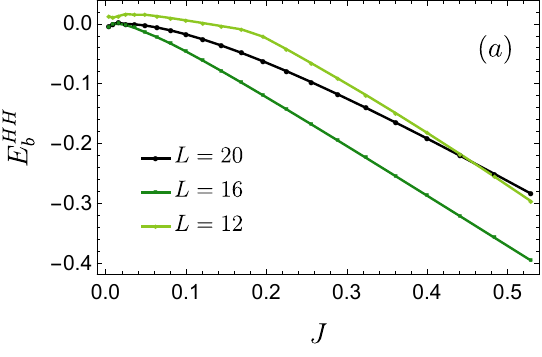}
\includegraphics[width=\columnwidth]{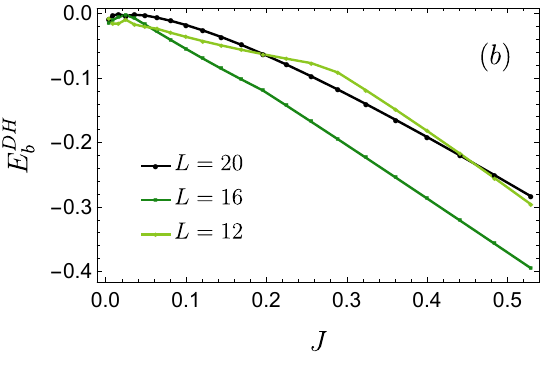}
\caption{Binding energy versus the isotropic superexchange $J=J_{\parallel, S/\eta}=J_{\perp, S/\eta}$ for (a)~ a holon-holon pair and for (b)~ a    doublon-holon pair at $L=12,16,20$ and $U=10$. 
Binding is present for a large regime of $J$ and increases with $J$, since breaking of spin and $\eta$ fluctuations is more and more energetically penalized, leading to the localization of charged pairs.}
\label{fig:fig2}
\end{figure}

The tendency towards binding between charged particles can be quantified by the binding energy~\cite{dagotto1994,troyer1996,tohyama2006,Zala2013PRL}, which for the chemically doped state is expresses as $E_b^{HH}= E_0^{HH}+E_0^{0} - 2 E_0^{H}$ and for the photo-doped states by $E_b^{DH}= E_0^{DH} + E_0^{0} -E_0^{D} -E_0^{H}$, where $E_0^{HH},  E_0^{DH}, E_0^{H}, E_0^{D},  E_0^{0},$ refer to the ground state energy of $H_{\text{eff}}$ in the sector with two holons, one doublon-holon pair,  one holon,  one doublon, and only singly occupied sites, respectively. The criterion for the formation of a bound pair is a negative binding
energy, $E_b^{DH}, E_b^{HH}<0$. In all of the results below, we will measure the binding energy in the units of $t_\parallel$. We should stress that the analysis below suffers from a particular finite size effect: while $E_0^{0}, E_0^{HH},E_0^{DH}$ are energies of ground states at zero momentum, ground state for the sector with a single holon/doublon appear at a finite momentum $k_x=\pi/2$, which might not be present for a given system size $L$. In any case, we use the solution with lowest energy.  To validate the robustness of our conclusion despite finite size difficulties, we will further compare the binding energy with real-space correlators to support  the binding. 

\begin{figure}[t!]
\includegraphics[width=\columnwidth]{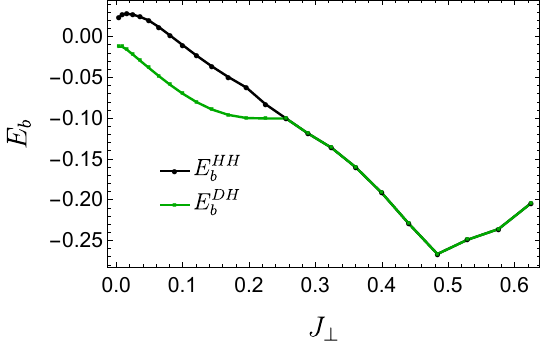}
\caption{Binding energy in the anisotropic system with varying perpendicular superexchange $J_{\perp}$ and fixed parallel superexhange $J_{\parallel}=0.4$} for a doublon-holon pair~(green) and for two holes~(black lines). Initially, binding energy grows with increasing $J_{\perp}=\frac{4 t_{\perp}^2}{U}$, but then the trend is reversed due to the competition of charge delocalization, promoted by the increase in $t_{\perp}$, and localization aided by large $J_{\perp}$. Parameters: $L=20$, $t_\parallel=1$, $U=10$.
\label{fig:fig3}
\end{figure}

We start the analysis with the isotropic case ($J=J_{\parallel, S/\eta}=J_{\perp, S/\eta}$) and show that a finite binding is 
% for the largest system considered ($N=20$) 
present for a large parameter regime, $J \gtrsim 0.15$, and it increases as a function of superexchange coupling, see Fig.~\ref{fig:fig2}. At large $J$, the magnitude of binding between a DH pair and two holons become identical, while for small values of $J$ the binding energies for the two kinds of exciton differ; however the difference is getting smaller with increasing system size. The increased binding in the limit of large superexchange $J$ comes from the fact that the breaking of spin fluctuations is more and more energetically penalized, leading to localization of the charged pair. In Fig.~\ref{fig:fig3} we show the general trend with increasing $J$, however, at larger $J$ higher order correction not discussed here could start to play a role.

Recent theoretical and experimental studies on the binding of two holons showed that by using external electric field~\cite{bohrdt2022,Hirthe2023} one can create highly anisotropic superexchange couplings, which can be favourable for binding. Here, we would like to explore this perspective by tuning the anisotropy to understand which parameter regime is favourable for both types of bindings. Fig.~\ref{fig:fig3} shows the binding energy as a function of $J_{\perp}$, keeping the parallel exchange parameter fixed $J_{\parallel}=0.4$. Initially, binding energy grows with increasing $J_{\perp}=\frac{4 t_{\perp}^2}{U}$, but then the trend is reversed due to the competition of charge delocalization, promoted by the increase in $t_{\perp}$, and localization aided by large $J_{\perp}$.
This suggests that ideally, we would like to decouple $t_\perp$ and $J_\perp$ to increase only the latter. In Section~\ref{floquet}, we will elucidate the methodology employed to accomplish this objective through Floquet engineering.

In the case of holon-doublon pairing, one might naïvely expect that including both spin and $\eta$ pairing would lead to cooperative enhancement of pairing in comparison to chemically doped situations. However, Fig.~\ref{fig:fig3} shows that the binding is comparable in the whole parameter range, meaning that spin and $\eta$ pairing neither cooperate nor compete, and such unexpected outcome is the first important message of this paper. In App.~\ref{eta} we compare in further details the calculation of DH binding energy including and excluding the $H_\eta^{\parallel(\perp)}$ terms and find that the inclusion of $\eta$ terms actually slightly reduces the binding. Additionally, we notice that the binding energy for the two different charged pairs becomes equal for $J_{\perp} \gtrsim 0.25$.

A useful starting point to understand the observation is to analyze the ground state of a 2$\times$2 cluster for the anisotropic case, $J_\perp>J_\parallel$, such that the DH or HH pairs are present in the same rung to reduce the energy lost by the breaking of the spin bonds~\cite{dagotto1994,troyer1996}. For simplicity, we have neglected the hopping terms, which decrease the number of basis states in the ground state manifold; see Appendix~\ref{appcluster} for details. Under this assumption, the ground state manifold for two holons consists of two states, $|\frac{ 0 \uparrow }{ 0 \downarrow} \rangle$ and $|\frac{ 0 \downarrow }{ 0 \uparrow} \rangle$,
where the top row represent the spin configuration in the upper chain and bottom in the lower chain. The ground state energy of this manifold is $\tilde{E}_{0}^{HH}=-J_{\perp,S}$. For the case of one DH pair, we work in the manifold with four states, 
$| \frac{0 \uparrow}{\updownarrow \downarrow} \rangle$, $\mid \frac{0 \downarrow}{\updownarrow \uparrow} \rangle$, $\mid \frac{\updownarrow \uparrow}{ 0 \downarrow }\rangle$ and $\mid \frac{\updownarrow \downarrow}{ 0 \uparrow} \rangle$. In equilibrium, the ground state energy is $\tilde{E}_{0}^{DH}=-(J_{\perp,S} + J_{\perp,\eta})/2$ and equals the binding energy for two holes, however, this is not anymore the case as we drive the systems, see below. Note that the calculation at this level of approximations thus cannot discern between different role of spin and $\eta$ fluctuations, observed in numerical results. We have neglected the hopping terms to obtain analytical estimations, which also differ between holons and doublons. While holon (as a quasiparticle) has a negative hopping sign, doublon has a positive one since by moving a doublon, we always hop over one fermion. At least in 2D systems, this can crucially influence the symmetry of the bound states, as pointed out in Refs.~\onlinecite{tohyama2006,Zala2013PRL,shinjo2021}. Although the above analysis is not exact, it explains the qualitative features we observe in the numerical analysis.

\subsection{Charge and spin correlators}
An alternative way to characterize the properties of bound states are real space correlators.
% To characterize the localization properties of bound states, we use two types of correlators. 
The charge correlators determine the relative spatial distribution of charged carriers. For the DH and HH pair, they are defined as 
$\chi_c^{DH}(\boldsymbol{\delta}{r}_j)\equiv \frac{1}{N} \sum_i \langle n_{i+j,\uparrow}n_{i+j,\downarrow} \bar{n}_{i,\uparrow} \bar{n}_{i,\downarrow} \rangle$ and 
$\chi_c^{HH}(\boldsymbol{\delta}\textbf{r}_j) \equiv \frac{1}{2N} \sum_i \langle \bar{n}_{i+j,\uparrow}\bar{n}_{i+j,\downarrow} \bar{n}_{i,\uparrow}\bar{ n}_{i,\downarrow} \rangle$, respectively. Furthermore, to capture the relative alignment of the up and down spins we define the spin correlator $\chi_s(\boldsymbol{\delta}\textbf{r}_j) \equiv \frac{1}{N} \sum_i \langle S_{i+r}^z S_i^z \rangle$.
\begin{figure*}[t!]
\centering {\ing[width=0.4\linewidth]{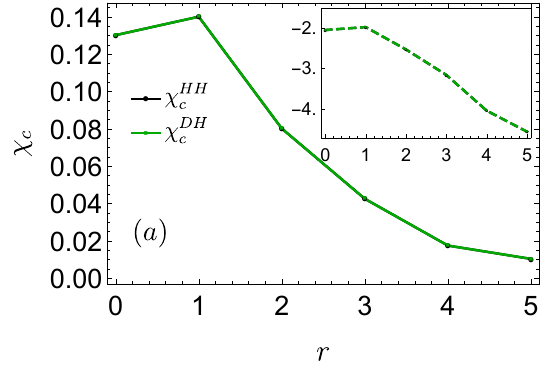}
\ing[width=0.4\linewidth]{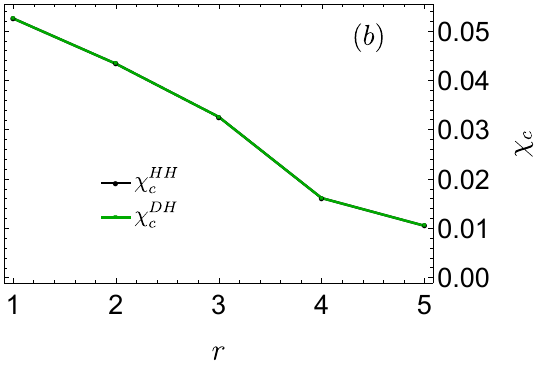}}
\centering {\ing[width=0.4\linewidth]{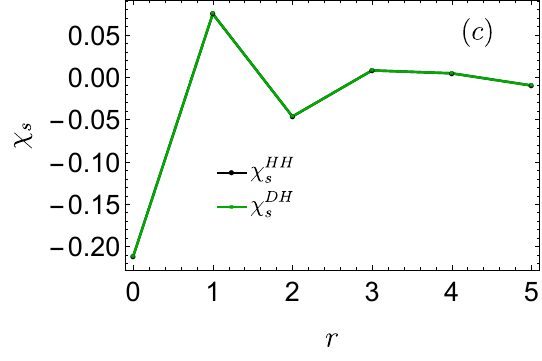}}
\centering {\ing[width=0.4\linewidth]{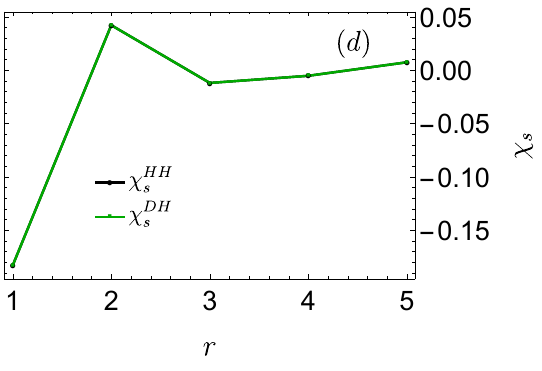}}
\caption{(a) Inter-chain $\chi_{c}(\{1,r\})$ and (b) intra-chain charge correlation $\chi_{c}(\{0,r\})$ reveal that for $J_{\perp}=J_{\parallel}=0.4$ charged particles prefer to sit in different chains along the diagonal of the rung. Supposedly exponentially decaying charge correlations (inferred from the inset showing the same data on the log scale) are consistent with a bound nature of states considered. (c) Inter-chain $\chi_{S}(\{1,r\})$ and (d) intra-chain spin correlation $\chi_{S}(\{0,r\})$ for doublon-holon and holon-holon pairs reveal antiferromagnetic local spin fluctuations which mediate the binding. Parameters: $L=20$, $J_{\perp}=J_{\parallel}=0.4$.}
\label{fig:fig4}
\end{figure*}

We show the charge correlator (top) and spin correlator (bottom) between the chains (left) and along the chain (right) in Fig.~\ref{fig:fig4}. 
For the isotropic case $J_{\perp}=J_{\parallel}=0.4$ shown, not only the binding energies $E_b^{HH} = E_b^{DH}$, c.f. Fig.~\ref{fig:fig3}, but also the correlators are indistinguishable for the DH and HH pairs. 
The rapidly decaying charge correlations for charges positioned in different legs, Fig.~\ref{fig:fig4}(a) are in agreement with the state being bound. It is worth noting that the binding along the same leg is considerably weaker, Fig.~\ref{fig:fig4}(b). 
The spin correlator, Fig.~\ref{fig:fig4}(c,d), confirms anti-ferromagnetic local fluctuations, which mediate the binding. While hopping terms tend to delocalise the charged pair, this causes a disturbance in the antiferromagnetic spin pattern. The competition between the two processes yields the existence of non-trivial bound states, with, for example, the strongest occupation of charged particles along the diagonal and not along the rung bond, as would be the case in the absence of hopping terms \cite{PhysRevB.55.6504}. For the DH pair, we have also evaluated the real-space superconducting correlator~(not shown), which shows staggered long-range correlations characteristic for $\eta$-pairing state and consistent with recent time-dependent ED study~\cite{ueda2023}.

\subsection{Enhanced Binding by Floquet Drive}
\label{floquet}
Increasing the exchange coupling $J_{\parallel(\perp)}=\frac{4t_{\parallel(\perp)}^2}{U}$ by reducing the interaction $U$ is one way to achieve the enhanced binding (Fig.~\ref{fig:fig2}), however, this approach has limitations since we must remain in the perturbative regime $4t_{\parallel(\perp)}\ll U$. 
Alternatively, tuning the rung hopping $t_\perp$, resulting in the anisotropy $J_{\perp}\neq J_{\parallel}$, in some cases leads to enhanced binding energy as well (Fig.~\ref{fig:fig3}), however, the enhancement is limited by the increase in $t_{\perp}$ which tend to delocalize the charges. Here we would like to explore whether manipulating the hopping and the exchange coupling independently could even further increase the stability of bounds states. Previously, this was achieved by introducing a potential term in one of the rungs~\cite{Hirthe2023}. We further harness this idea by introducing a time-periodic field and explore the Floquet engineering of the effective model to enhance the stability of bounds states.

We drive the system with an external electric field applied along the rung and introduced via Peierls substitution to the hopping term, $t_{ij}(\tau) = t_{\perp} e^{i \boldsymbol{A}(\tau).\boldsymbol{r}_{ij}}$, between nearest neighbor sites $i$ and $j$. Here, $t_{\perp}$ is the equilibrium rung hopping parameter, $\boldsymbol{A}(\tau)$ is the vector potential and $\boldsymbol{r}_{ij}$ is the vector from $j$ to the $i$ site. The charge of the particles and distance between nearest neighbours is considered unity; hence the electric field is related to the vector potential by $\boldsymbol{E}(\tau) = -\partial \boldsymbol{A}(\tau)/\partial t$. While we formulate our Floquet theory in terms of the time-dependent vector potential modifying the hopping, the equivalent formulation in terms of the electric field modifying the potential could be more easily implemented in cold atoms experiment \cite{sandholzer19}. A further option is a periodic drive of the hopping integral, which we discuss in Sec.~\ref{coldatom}. For field in the direction of the rung $\boldsymbol{A}(\tau) =\{0, A \sin(\omega \tau)\}$, the hopping and exchange couplings along the chain ($t_\parallel$, $J_\parallel$) remain the same as in the equilibrium.
Let us write the interaction strength as $U=l_0 \omega+ \Delta U$, where $U_0 = l_0 \omega$ is the closest approximation to $U$ in multiples of $\omega$ and $\Delta U$ act as the detuning. The Fourier components are then expressed as,
\begin{align}\label{Eq:Floquet}
t_{ij}(\tau) &= t_{\perp} \sum\limits_{l} A_{\boldsymbol{r_{ij}}}^{(l)} e^{i l \omega \tau} 
%= t_{\perp} e^{-i U_0 t}  \sum\limits_l B_{ij}^{(l)} e^{i l \omega t} \notag \\ 
\notag \\
A_{\boldsymbol{r}_{ij}}^{(l)} &= \frac{1}{T} \int_{-T/2}^{T/2}  e^{i \boldsymbol{r}_{ij} \cdot \boldsymbol{A}(\tau)} e^{-i l \omega \tau} d\tau =  \mathcal{J}_{l} (\boldsymbol{r}_{ij} \cdot \boldsymbol{A}),
\end{align}
where
$\mathcal{J}_{l}$ is the Bessel function of order $l$.  Here, we have used the relation $\mathcal{J}_{l}(x)=\frac{1}{2 T} \int_{-T}^{T} e^{i(x~\sin \omega \tau- l \omega \tau)} d\tau$.

\begin{figure}[t!]
\centering {\ing[width=0.9\linewidth]{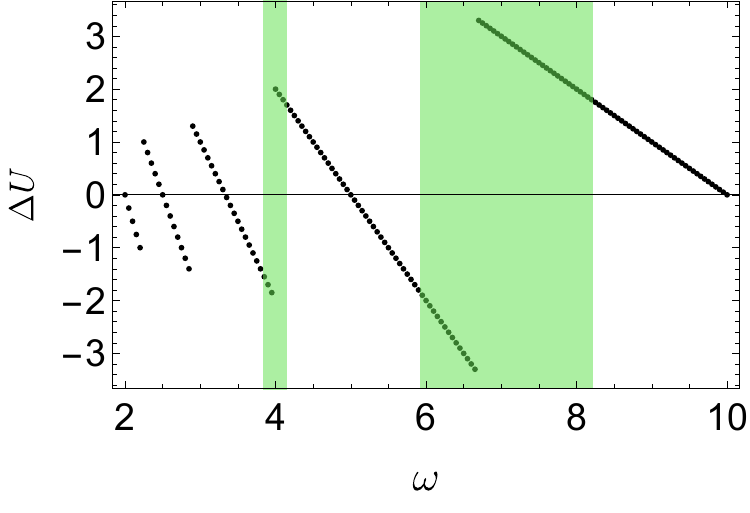}}
\caption{Detuning parameter $\Delta U$ dependence on the driving frequency $\omega$ with regions of validity for the off-resonant driving $U, \omega, |\Delta U| \gg t_{\perp}$. As off-resonant we consider situations with $|\Delta U| > 1.5 t_{\perp}$, shaded in green. The interaction parameter is $U$=10.}
\label{fig:fig5}
\end{figure}

\begin{figure}[t!]
\centering {\ing[width=0.8\linewidth]{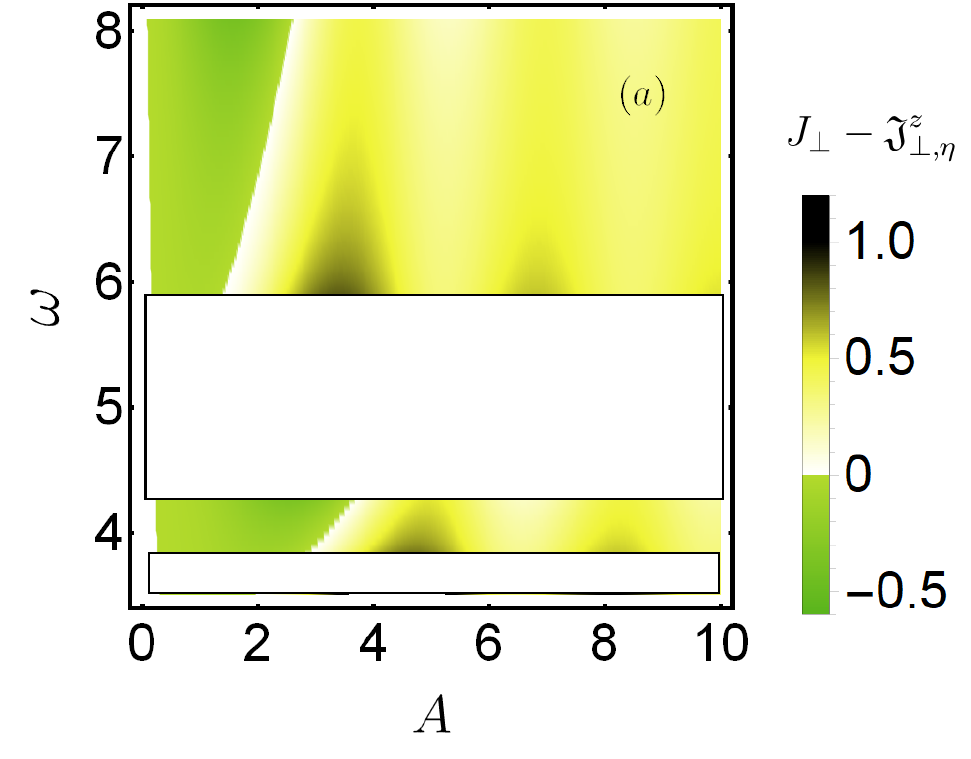}}\\
\centering {\ing[width=0.8\linewidth]{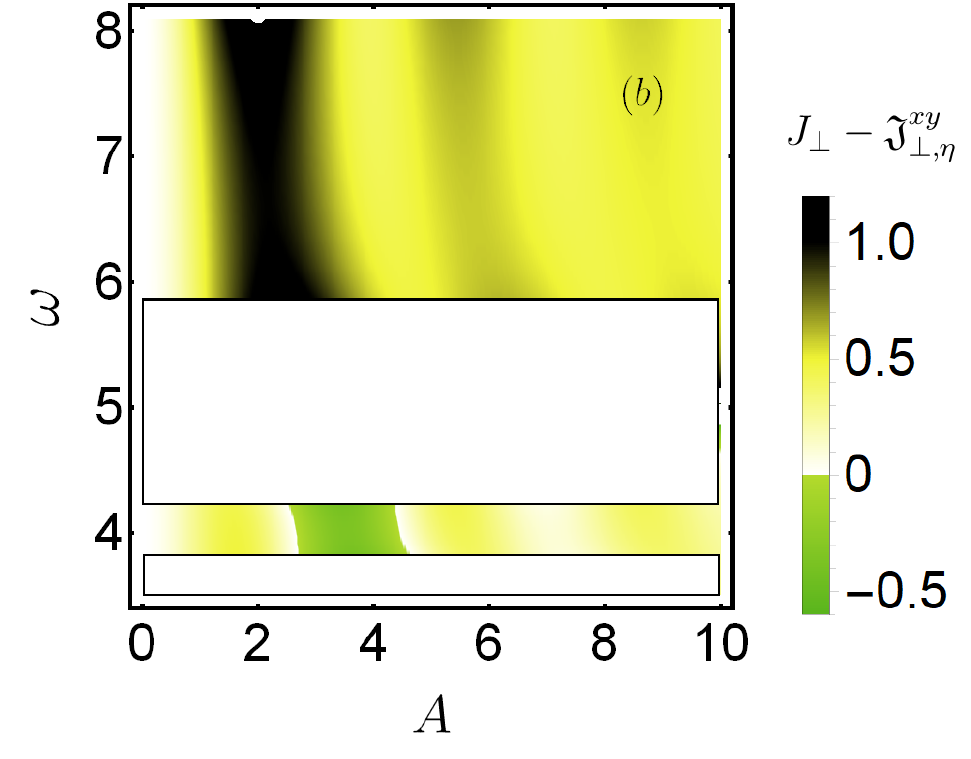}}
\caption{Variation of the exchange parameters for off-resonant driving with respect to the equilibrium value for~(a)  the spin channel $J_\perp - \mathfrak{J}_{\perp,S}$ and~(b) the $\eta$ exchange channel $J_\perp - \mathfrak{J}_{\perp,\eta}^{xy}$  with driving frequency $\omega$ and the driving strength $A$. These plots corresponds to $U=10$, $t_{\perp}=t_{\perp}=1$. White boxes in the plot indicates the regions that do not obey the conditions for  the off-resonant condition, $\Delta U > 1.5 t_\parallel$. }
\label{fig:fig9}
\end{figure}

Depending on the value of $\Delta U$ there are different regimes; far-from-resonant driving with~$U, \omega, |\Delta U| \gg t_{\perp}$ and near-resonant driving with $U, \omega, \gg t_{\perp}, |\Delta U|$, which can be treated with appropriate high-frequency expansions \cite{Bukov2016,Bukov2015,Mentink2015,murakami2023suppression}. We will consider only the off-resonant driving, in Fig.~\ref{fig:fig5} denoted by the green regions, as heating and doublon-holon recombination effects are suppressed in this regime, see Sec~\ref{sec:recom}. In Figs.~\ref{fig:fig5} and \ref{fig:fig9} we have used $\Delta U > 1.5~ t_{\perp}$ as condition for the off-resonant regime.

After performing the high-frequency expansion in the off-resonant regime and neglecting higher order terms, see App.~\ref{hamiloff}, we arrive
at the effective Floquet Hamiltonian $H_{\text{eff}}^{(F)}$ of form \eqref{eq::Heff}, where parameters along the chains equal to the equilibrium ones, while both hopping and exchange parameters in the rung (field) direction depend on the drive parameters and get modified as
\cite{Bukov2016,Bukov2015,Mentink2015,murakami2023suppression}
\begin{eqnarray}
\label{r2exparam}
\mathfrak{t}_{\perp}&=& t_{\perp} \mathcal{J}_0(A)\\
\mathfrak{J}_{\perp, \eta}^{xy} &=& \sum_{l=-\infty}^{\infty} \frac{ 4 t_{\perp}^2}{(U-l \omega)}  A_{\boldsymbol{e}_y}^{(l)} A_{-\boldsymbol{e}_y}^{(l)} \label{JxyFloq} \\ 
\mathfrak{J}_{\perp, \eta}^{z} &=&  \mathfrak{J}_{\perp, S} = \sum_{l=-\infty}^{\infty}  \frac{4 t_{\perp}^2}{(U-l \omega)} \mid A_{\boldsymbol{e}_y}^{(l)} \mid^2. \label{JzFloq}
\end{eqnarray} 

Note that in this case the isotropy between the $\eta-$exchange parameters is broken $\mathfrak{J}_{\perp, \eta}^{z} \neq \mathfrak{J}_{\perp, \eta}^{xy},$ unlike in equilibrium. Applying a periodic drive thus gives the freedom to vary hopping and different exchange parameters independently. Moreover, unlike the mixed-dimensional approach~\cite{bohrdt2022, bourgund2023formation} which allows tuning of only the exchange parameters ($J_{\perp}^z$,~$J_{\perp}^{xy}$), the Floquet drive allows one to additionally modify the hopping parameter ($t_{\perp}$) as a function of field strength. In Fig.~\ref{fig:fig9}, we show the variation of (a) $J_\perp - \mathfrak{J}_{\perp,S}$ and (b) $J_\perp - \mathfrak{J}_{\perp,\eta}^{xy}$ as a function of the field strength $A$ and frequency $\omega$.
White boxes in the plot correspond to areas that do not obey the off-resonant limit, $U, \omega, |\Delta U| \gg t_{\perp}$.
Regions with $J_\perp < \mathfrak{J}_{\perp,S}$ (stronger than equilibrium spin exchange coupling) are candidate regions in which enhanced exchange couplings could mediate stronger binding between the charged particles. The corresponding role of $\eta$ exchange coupling is less clear, but we plot in Fig.~\ref{fig:fig9}(b), $J_\perp < \mathfrak{J}_{\perp,\eta}^{xy}$ for completeness.

\begin{figure*}[t!]
\centering {\ing[width=0.4\linewidth]{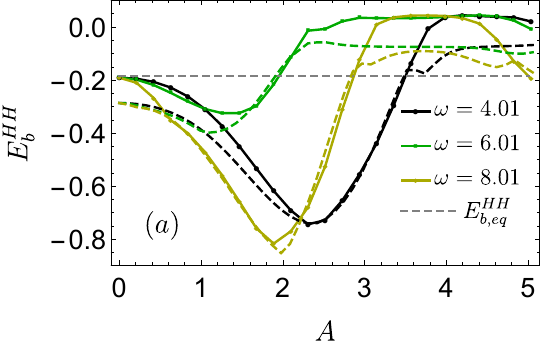}}
\centering {\ing[width=0.4\linewidth]{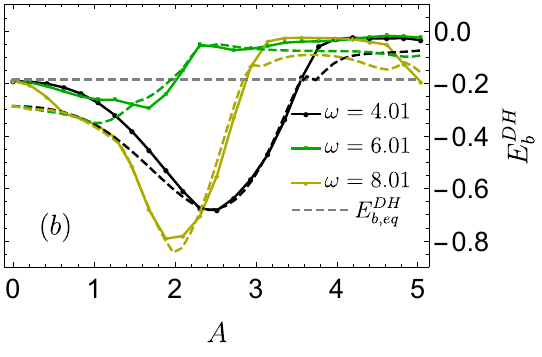}}
\centering {\ing[width=0.4\linewidth]{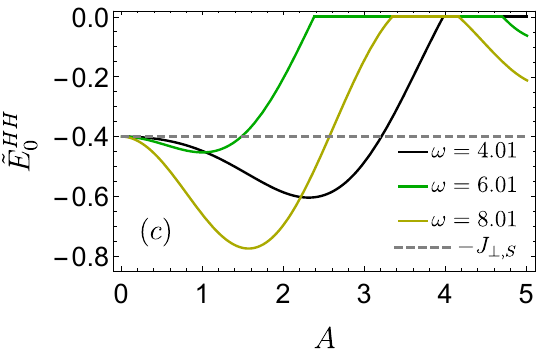}}
\centering {\ing[width=0.4\linewidth]{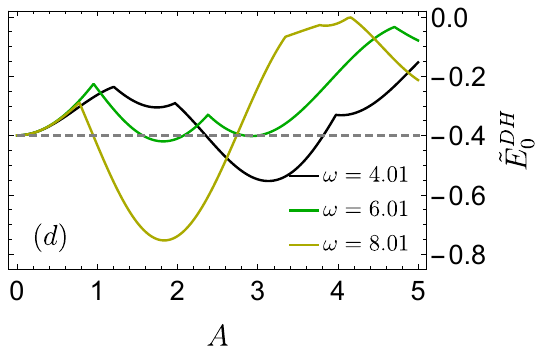}}
\hspace{-0.3in} 
\centering {\ing[width=0.63\linewidth]{{fig6en}.pdf}}
\caption{Binding energy in a Floquet driven setup for (a)~holon-holon and (b)~doublon-holon pair as a function of driving amplitude $A$ at various driving frequencies $\omega=4.01,6.01,8.01$. The dashed lines indicates the behaviour of binding energy at same parameters but for smaller system $L=16$. (c-d) Variation of the estimated ground state energy for the 2$\times$2 cluster including one holon-holon pair (c) and one doublon-holon pair (d). The horizontal lines mark the equilibrium~(non-driven) value. (e) Variation of Floquet modified hopping  $\mathfrak{t}_{\perp}$ as a function of field strenght $A$.
%Variation of the estimated binding energy from the 2$\times$2 cluster for the HH pair~$-\mathfrak{J}_{\perp,s}$~(c) and the DH pair~$E^{DH}_{\text{clust}}$~(d). The horizontal lines mark the equilibrium~(non-driven) value.(d) Variation of Floquet modified hopping  $\mathfrak{t}_{\perp}$ as a function of field strenght $A$.
Parameters: $L=20$, $U=10$, $t_{\parallel}=t_{\perp}=1$.} 
\label{fig:fig10}
\end{figure*}

Now, we will check how well such predictions work by analyzing the binding energy of charge carriers within the effective Floquet Hamiltonian $H_{\text{eff}}^{(F)}$. We will focus on the field amplitude $A$ dependence at three frequencies, $\omega=4.01, 6.01,8.01$.  
Figs.~\ref{fig:fig10}(a,b) shows the binding energy for the HH and DH pair, respectively. In Figs.~\ref{fig:fig10}(c,d), we show for comparison the HH and DH ground state energies estimated from the analytical calculations on 2$\times$2 clusters with a HH or DH pair, introduced in Sec.~\ref{sec::results}A and explained in App.~\ref{appcluster}. Figs.~\ref{fig:fig10}(c) plots the HH expression, $\tilde{E}_0^{HH}=-\mathfrak{J}_{\perp,S}$ for $\mathfrak{J}_{\perp,S}>0$ and $\tilde{E}_0^{HH}=0$ for $\mathfrak{J}_{\perp,S}<0$. Figs.~\ref{fig:fig10}(d) gives the DH ground state energy dependence $\tilde{E}_0^{DH}(A)$. In equilibrium or at weak driving strength $A$, the ground state energy is given by $\tilde{E}_0^{DH}=-(\mathfrak{J}_{\perp,\eta}^{xy}+\mathfrak{J}_{\perp,S})/2$, but for larger driving other states can become the ground state. Indeed, the binding energy for the HH and DH pair is increased in the similar range as  predicted by 2$\times$2 cluster estimations. Also note from Fig.~\ref{fig:fig10}(e) that the hopping strength $\mathfrak{t}_{\perp}$ is substantially reduced in this regime of $A$ which is further favourable for binding. Overall, the conclusion is that the Floquet driving leads to a substantial modification of the pair binding energies, which can be enhanced to more than three times compared to the equilibrium values.

\begin{figure*}[t!]
\centering {\ing[width=0.4\linewidth]{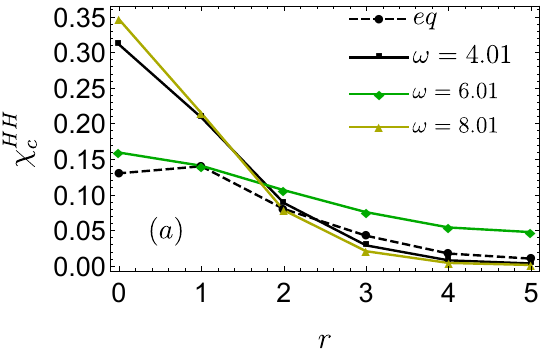}}
\centering {\ing[width=0.4\linewidth]{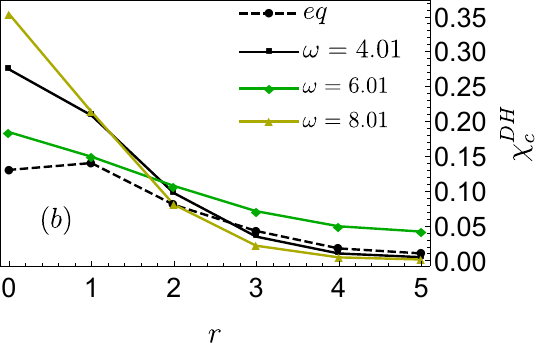}}
\centering {\ing[width=0.4\linewidth]{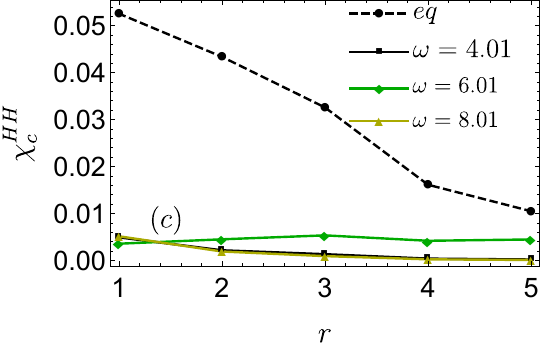}}
\centering {\ing[width=0.4\linewidth]{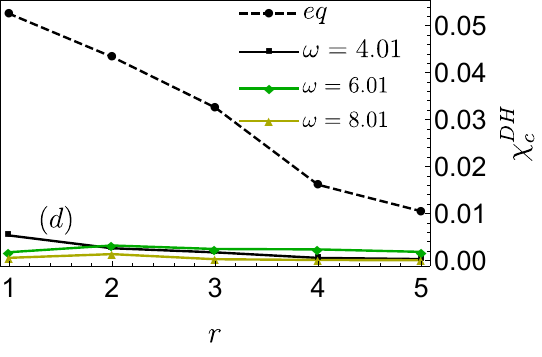}}
\caption{Inter-chain change correlators $\chi_{c}(\{1,r\})$ for (a) holon-holon and (b) doublon-holon pair for different driving frequencies $\omega=4.01,6.01,8.01$ and in equilibrium~(eq). (c,d) Same for the intra-chain correlator. Parameters: $t_{\parallel}=t_{\perp}=1$, $U=10$, $A=1.98$ and $L=20$.}
\label{fig:fig11}
\end{figure*}

To further ensure that the enhanced binding energies indeed lead to a more bound pair we consider real-space HH and DH correlators. In Fig.~\ref{fig:fig11}, we compare the charge correlators for the system under the Floquet drive to its equilibrium counterpart. The inter-chain correlators $\chi_{c}(\{1,r\})$ show that in the regions of the enhanced bindings (e.g. A=1.98), the charge carriers are indeed residing at substantially closer distances both for HH and DH pair, see Fig.~\ref{fig:fig11}(a) and (b). Furthermore, the intra-chain correlators $\chi_{c}(\{0,r\})$ shows a strong reduction with respect to the equilibrium value, which means that the two charge carriers will tend to form pairs in the opposite chain along the rung. These results show that the Floquet driving can substantially increase the tendency of charge carriers to form bound pairs. 

Based on the above analysis, we find that Floquet engineering can result in significantly enhanced binding energies due to the reduction of the hopping integral and the enhancement of spin and $\eta$ fluctuations. At least in the two-leg ladder system, enhancing HH binding appears in a broader regime of drive parameters $(\omega, A)$ since it only requires increased $J_{\perp,S}$. To increase DH binding, we preferentially want to increase both $J_{\perp,S}$ and $J_{\perp,\eta}^{xy}$, which seems to be exclusive in many cases. However, the enhanced DH binding is still present in a broad parameter regime as the Floquet driving decreases the effective hopping integral, leading to a higher tendency for excitonic bindings. Since the Floquet driving increases binding and thus decreases the localization length of the bound states, it also leads to a reduction of the finite size effect in our calculation as shown by the dashed lines corresponding to the plots of binding energy for $L=16$ in Fig.~\ref{fig:fig10}(a) and (b).

\section{Results for two dimensional square lattice}\label{sec:2D}

Now, we analyze how these findings are translated to 2D systems. We will consider 2D lattices with periodic boundary conditions for which the $x$ and $y$ direction are equivalent; therefore, the $\perp$ and $\parallel$ terms in the effective model $H_{\text{eff}}$, Eq.~\eqref{eq::Heff}, are equivalent. At least in equilibrium, i.e., without any Floquet engineering, all the exchange couplings are the same and equal $J_{\parallel,S}=J_{\parallel,\eta}=J_{\perp,S}=J_{\perp,\eta}=\frac{4t^2}{U} \equiv J$.

\begin{figure}[b!]
\includegraphics[width=\columnwidth]{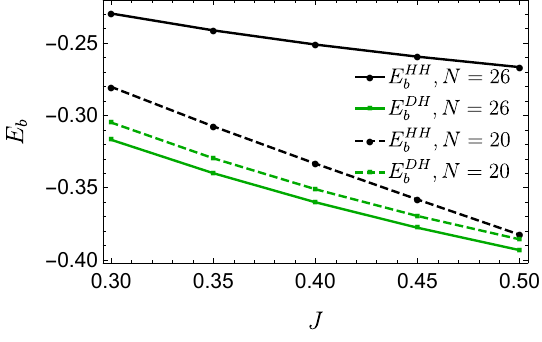}
\caption{Binding energy for a doublon-holon pair (black) and holon-holon pair (green) as a function of $J$ for $N=20,26$.}
\label{fig:fig13}
\end{figure}

The question of HH and DH binding in 2D has already been addressed theoretically~\cite{tohyama2006,Takahashi2002,Zala2013PRL,lenarcic13,shinjo2021,Jaklic2000,dagotto1994}, as of relevance for the high-temperature superconductivity in effectively two dimensional doped cuprates \cite{dagotto1994} and photo-doped charge transfer and Mott insulating materials \cite{Okamoto2010,Okamoto2011,Okamoto2019SciAdv,Mehio2023}. 
It has been recently experimentally confirmed that in photo-doped charge transfer and Mott insulating materials, bound states of doublon-holon pairs, i.e., Hubbard excitons, indeed appear as metastable states that form after the pump pulse. They can be detected via a transfer of spectral weight from the Drude peak to a finite frequency Lorenzian \cite{Mehio2023} and in the third order optical response~\cite{Okamoto2019SciAdv}. Another indirect clue for the presence of doublon-holon excitons comes from an exponential functional form of the mid-infrared peak decay in photo-doped materials \cite{Okamoto2010,Okamoto2011,sahota2019}, reflecting that the recombination process of doublon-holon pair happens from a bound state. The recombination rate is exponentially suppressed in the number of bosonic excitations that are emitted in the process, in order to bridge the Mott gap \cite{Zala2013PRL,lenarcic13}. Furthermore, it was shown that the non-local Coulomb interaction can cooperate with the super-exchange coupling assisted excitons formation~\cite{Zala2013PRL,Bittner2020,Mehio2023}.

Here we revisit the equilibrium analysis and compare the HH and DH binding. As opposed to the previous studies of DH Hubbard excitons \cite{tohyama2006,Takahashi2002,Zala2013PRL,lenarcic13,Mehio2023}, we include here the effect of $\eta$ fluctuations, by keeping the $\eta$ term $H_{\eta}$ in the effective Hamiltonian $H_{\text{eff}}$, Eq.~\eqref{eq::Heff}, that we diagonalize using Krylov technique in the sector with one holon-holon and one doublon-holon pair on system of size up to $N=26$ sites. The 2D lattice geometries corresponding to $L=20$ and $L=26$ have been presented in Refs.~\onlinecite{Jaklic2000,Prelovsek_springer}.

\subsection{Equilibrium results}

\begin{figure}[t!]
\includegraphics[width=0.49\columnwidth]{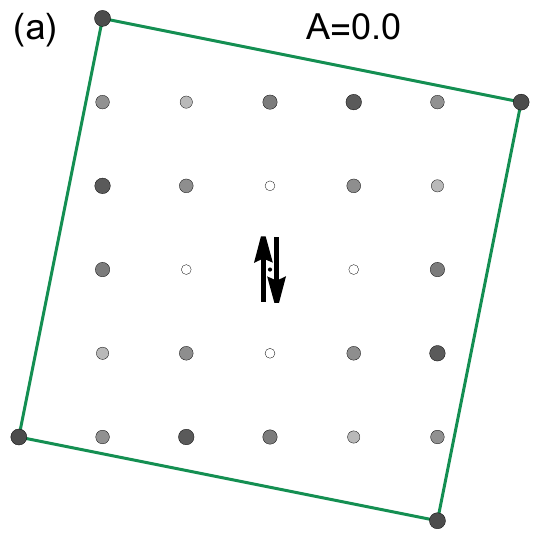}
\includegraphics[width=0.49\columnwidth]{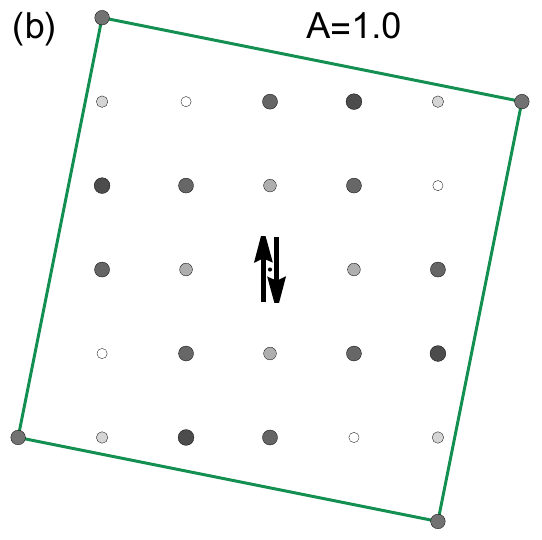}\\
\includegraphics[width=0.49\columnwidth]{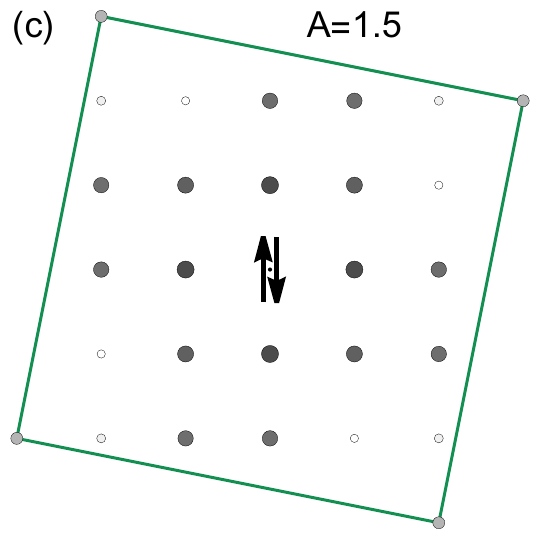}
\includegraphics[width=0.49\columnwidth]{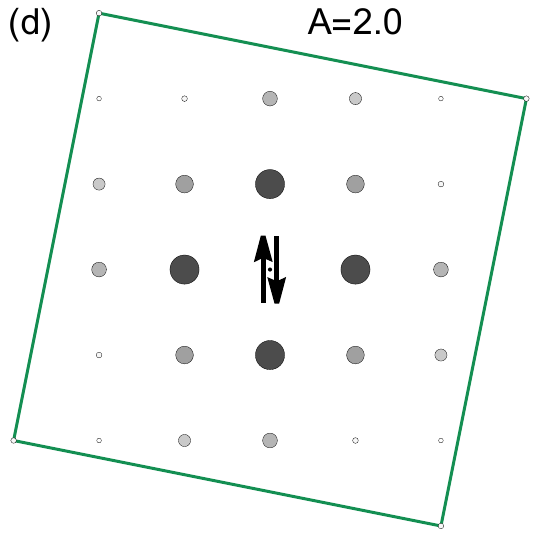}

\caption{Charge correlator $\chi_c^{DH}$ showing the distribution of holon relative to the doublon (in origin) for the doublon-holon exciton state (a) in the equilibrium at $J=0.4$ ($A=0$) and (b-d) for the Floquet driving with $A=1.0,1.5,2.0$, $\omega=6.01$ and $U=10$. The size of the dots is proportional to the correlation on this site and suggests the presence of binding. Periodic boundary condition on considered $N=26$ sites lattice are denoted by the square.}
\label{fig:HDdistr}
\end{figure}

\begin{figure}[t!]
\includegraphics[width=0.49\columnwidth]{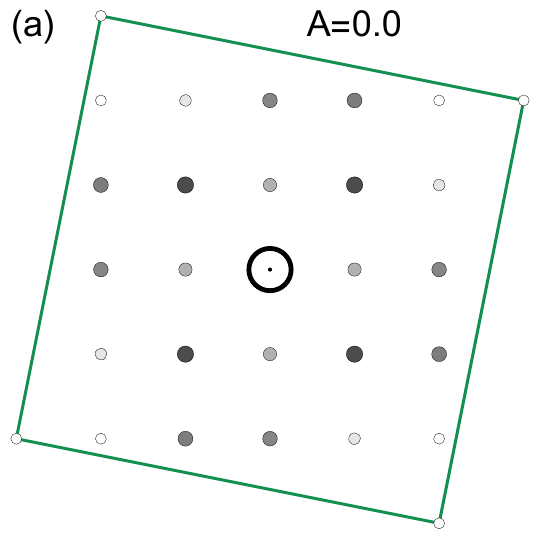}
\includegraphics[width=0.49\columnwidth]{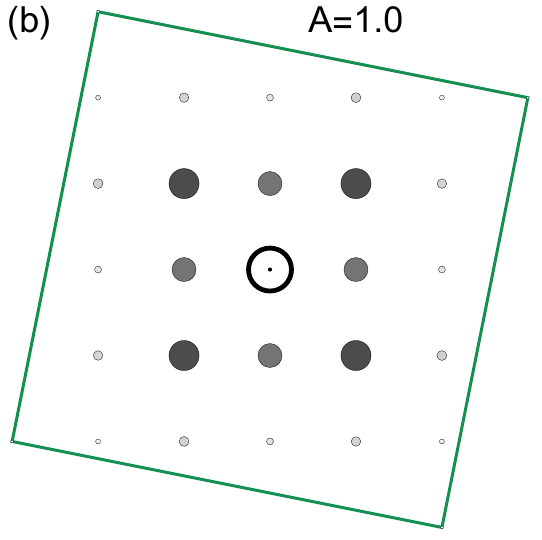}\\
\includegraphics[width=0.49\columnwidth]{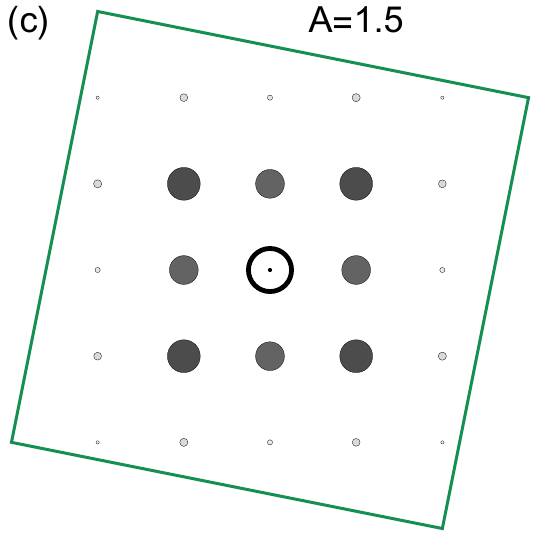}
\includegraphics[width=0.49\columnwidth]{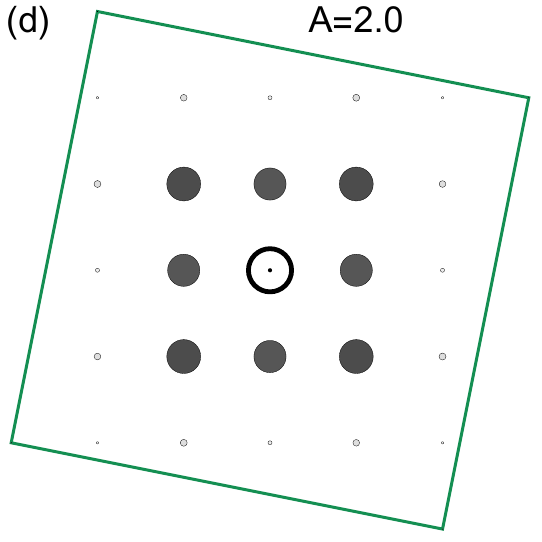}
\caption{Charge correlator $\chi_c^{HH}$, showing the distribution of the first holon relative to the second holon (in the origin) for the holon-holon bound state (a) in the equilibrium at $J=0.4$ ($A=0$), (b-d) for the Floquet driving with $A=1.0,1.5,2.0$, $\omega=6.01$ and $U=10$. Size of the dots is proportional to the correlation on this site and suggests the presence of binding. Periodic boundary condition on considered $N=26$ sites lattice are denoted by the square.}
\label{fig:HHdistr}
\end{figure}

In Fig.~\ref{fig:fig13}, we show binding energies for holon-holon and doublon-holon pair as a function of $J$ on systems of size $N=20,26$ sites. To minimize the finite size effects on lattices that do not have the $\boldsymbol{k}=(\pi/2,\pi/2)$ wave vector (which is the case for $N=20,26$), we use for $H_0^{H}$ and $H_0^{D}$ the fit obtained from $N=16,32$ in Ref.~\onlinecite{Leung1995}. As in the two-leg ladder, increasing the exchange coupling $J$ enhances the binding energy, since breaking the short-range antiferromagnetic spin fluctuations is energetically penalized, leading to localization of the pair. Our results suggest that in 2D, including the $\eta$ term $H_\eta$ slightly reduces the binding energy of DH exciton, see App.~\ref{eta}, which suggests that $\eta$ fluctuations do not help to stabilize the exciton. However, the $\eta$ contribution is not very significant so the Hubbard excitons are stable even if this term is taken into account. 
Unlike in the two-leg ladder setup, here, doublon-holon and holon-holon binding energies are always distinct, and this is true with or without including the $\eta$ fluctuations. We hypothesise that this can be traced back to the difference in phases in the hopping term for holon and doublon, Eq.~\eqref{eq:Dhop}, which is known to result in different wave function symmetry of the two bound states: while the holon-holon bound states has $d$-wave symmetry, the doublon-holon bound states has $s$-wave symmetry \cite{tohyama2006}. As some of us recently discussed in relation to experimental observation of intra-excitonic transitions \cite{Mehio2023}, there actually exist several DH Hubbard excitons with different symmetries. Here, however, we discuss only the lowest and most stable one. 

In Fig.~\ref{fig:HDdistr}(a), we plot the charge correlator $\chi_c^{DH}$, showing the distribution of holon relative to the doublon (in the origin) for the Hubbard exciton state at $J=0.4$. Results are in agreement with a bound state, however, we can see that the excitonic state is rather extended: the largest contribution to the excitonic wave function comes from holon and doublon being only the third nearest neighbors. 
In Fig.~\ref{fig:HHdistr}(a), we plot an equivalent charge correlator $\chi_c^{HH}$ for the holon-holon bound state, showing the distribution of one holon (in the origin) relative to the other one for $J=0.4$. Results are again in agreement with a bound state.

\subsection{Floquet engineering}
Lastly, we revisit the Floquet engineering of exchange coupling parameters from Sec.~\ref{floquet}. In the two-leg ladder, we considered the electric field along the rung with $\boldsymbol{A}(\tau)=\{0,A \sin(\omega t)\}$ which modulated only the exchange coupling along the rungs. Here, we consider the electric field applied along the diagonal, with $\boldsymbol{A}(\tau)=\{A \sin(\omega t),A \sin(\omega t)\}$. Then, the four-fold symmetry of the 2D lattice is preserved and exchange coupling between all the nearest neighbors are the same. We will again present results for the off-resonant regime with $U, \omega, |\Delta U| \gg t$, where different exchange terms are given by Eq.~\eqref{JxyFloq} and Eq.~\eqref{JzFloq}. 

\begin{figure}[b!]
\includegraphics[width=0.83\linewidth]{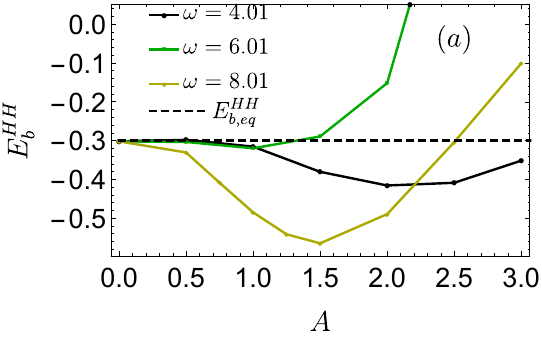}
\includegraphics[width=0.83\linewidth]{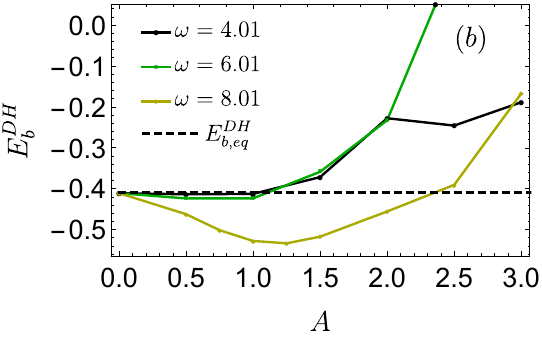}
\caption{Binding energy for a (a) holon-holon pair and (b) doublon-holon pair in the Floquet driven case as a function of driving amplitude $A$ at $\omega=4.01,6.01,8.01$ and $U=10$ for $N=26$. The horizontal line corresponds to the equilibrium value with $J=0.4$.}
\label{fig:EbFloq2D}
\end{figure}

In Fig.~\ref{fig:EbFloq2D}, we show the variation of the binding energy for holon-holon and doublon-holon pair as a function of $A$ for $\omega=4.01,6.01,8.01$ and $N=26$. We choose the same frequencies as in the corresponding study in the two-leg ladder. In this case, we use for $E_0^H,E_0^D$ the energies at the momentum with lowest g.s. energy for the given system size; the values of effective exchange parameters vs. the hopping are outside of the validity of fit from Ref.~\onlinecite{Leung1995} used in the static case. Different, yet for each case optimal choices thus lead to a small mismatch of equilibrium values between Fig.~\ref{fig:EbFloq2D} and Fig.~\ref{fig:fig13}, signaling one should consider the binding energies as qualitative not quantitative measure of the binding. Like in the ladder geometry, there is an extended regime of field strengths for which the binding is enhanced via the Floquet engineering of coupling parameters. This conclusion is supported also by the DH and HH correlators, which are in Fig.~\ref{fig:HDdistr}(b,c,d) and Fig.~\ref{fig:HHdistr}(b,c,d) shown for field strengths $A=1.0,1.5,2.0$. It is clearly visible that for a region of parameters $A$, the DH and HH pairs get more localized than at equilibrium. In the extreme case of large driving, e.g. $A=2.0$ for DH pair in Fig.~\ref{fig:HDdistr}, the DH pair resides at the nearest-neighbor sites in stark contrast to the equilibrium value. 

This result once again underlines that Floquet engineering is indeed a relevant approach that can significantly increase the localization of the bound states and should be a viable option that could be implemented in either solid state or cold atoms setups. For the latter, we draft our suggestion in Section~\ref{coldatom}. 

\section{Recombination of Hubbard excitons}\label{sec:recom}
While Floquet driving can increase the binding of charged pairs by modifying the effective hopping and exchange parameters, it also has an influence on the doublon-holon recombination rate that determines the timescale on which excitons are stable and observable. Namely, Floquet driving opens additional recombination channels and therefore it is important to understand how stable is the excitonic state in the driven case. 

Due to the large energy of the Mott gap that must be instantaneously released in the recombination process of a doublon-holon pair, the recombination timescale is much slower than the timescale for the intrinsic dynamics/relaxation of holons and doublons within each Hubbard band. Even though recombination appears in the Hubbard model within the hopping term of strength $t$, it is reasonable to treat it perturbatively: we have made this formal in Sec.~\ref{hamil} by canonically transforming the original model, Eq.~\eqref{sw1}, so that recombination term appears with a smaller prefactor (e.g., $t^2/U$) in the transformed one. Transformation yields the effective Hamiltonian $H_{\text{eff}}$, Eq.~\eqref{eq::Heff}, that preserves the number of doublon-holon pairs, using which we can elegantly estimate the metastable excitonic state. Moreover, the transformation also singles out the dominant recombination, now with a prefactor of the order $J$~\cite{lenarcic2012,Zala2013PRL},

\begin{equation}\label{eq::Hrc}
H_{\text{rc}} = \frac{J}{2} \sum_{(ijk)ss'} (1-n_{i,\bar{s}}) c_{is}^{\dagger} n_{ks'}c_{k\bar{s}'} \, \vec{\sigma}_{s \bar{s}'}\cdot \boldsymbol{S}_j,
\end{equation}
where $\bar{s}$ is spin opposite to $s$ and $i\neq k$ are nearest-neighbor sites to $j$.
In the static setup and at a low density of photoexcited DH pairs, the only viable recombination channel is via emission of spin excitations, which absorb the large energy of the Mott gap $\Delta$ as doublon and holon recombine across the gap. 
The recombination rate can be numerically estimated via Fermi's golden rule expression for the transition from the excitonic state $\ket{\psi_0^{DH}}$ into a highly excited state of spins $\ket{\psi_m^0}$ with energy $E_m^0$,
\begin{equation}
\Gamma = 2\pi \sum_m |\ave{\psi_m^0|H_{\text{rc}}|\psi_0^{DH}}|^2 \delta(E_m^0 - E_0^{DH}).
\end{equation}
In Refs.~\onlinecite{lenarcic2012,lenarcic13}, some of us showed that the recombination rate $\Gamma$ is roughy exponentially suppressed in the number of spin excitation $\Delta/J$ that are emitted in such a process. A more precise dependence on the model parameters was derived in Ref.~\onlinecite{PhysRevB.83.054508} by making a connection to the exciton-boson model, where exciton recombines by emitting some general boson excitations. In this case the decay rate can be approximated by
\begin{equation}\label{Eq:recombination}
\Gamma \approx g_{{\rm rc}}^2 e^{-\xi} \sqrt{\frac{2\pi}{\Delta \omega_0}}\exp\left[-\frac{\Delta}{\omega_0} \ln\left(\frac{\Delta}{e \xi \omega_0}\right)\right],
\end{equation}
where $g_{{\rm rc}}$ is the prefactor in the operator causing the recombination, $\xi$ is the dimensionless coupling strength between the exciton and bosons, and $\omega_0$ is the typical energy of bosons emitted. For coupling to spins, $\omega_0=J$, $\xi$ was fitted to be roughly $\xi\approx 3$, and $g_{{\rm rc}}$ was fitted to $g_{{\rm rc}} \approx J$, corresponding to the actual prefactor in $H_{\text{rc}}$, Eq.~\eqref{eq::Hrc}. 

In the case of Floquet driving with the original time-dependent Hubbard model, a strategy to estimate the recombination rate is to perform a time-dependent canonical transformation, which once again transforms out the recombination term (from the time-dependent model) at order $t$. Inspired by Ref.~\onlinecite{Kitamura2017}, we perform this transformation in App.~\ref{apprecomb} and obtain that in the lowest order of the high-frequency expansion, the dominant recombination term has the same structure as in the static case, but with a different prefactor,
\begin{align}\label{eq::HrcF}
H_{\text{rc}}^{(F)} = \frac{1}{2}\sum_{(ijk)ss'} \sum_{l} &
\frac{t_{ij} t_{jk} 
\mathcal{J}_{-l}(\boldsymbol{A}\cdot \boldsymbol{r}_{ji}) 
\mathcal{J}_{l}(\boldsymbol{A}\cdot \boldsymbol{r}_{jk})}{U-l \omega} \notag \\ 
&\times(1-n_{i,\bar{s}}) c_{is}^{\dagger} n_{ks'}c_{k\bar{s}'} \, \vec{\sigma}_{s \bar{s}'}\cdot \boldsymbol{S}_j ,
\end{align}

\begin{figure}[b!]
\hspace{-0.3in}\includegraphics[width=0.9\columnwidth]{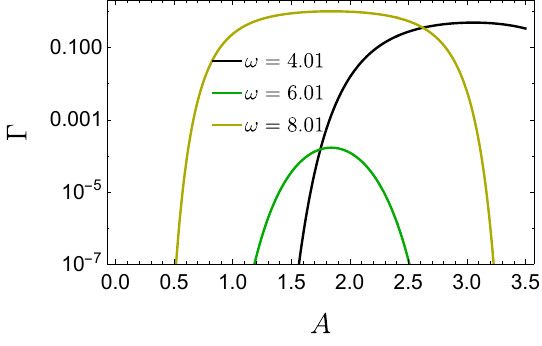}
\caption{Upper bound on the recombination rate $\Gamma$ as a function of the Floquet driving amplitude $A$, estimate from the expression \eqref{Eq:ffgr} for $\omega=4.01, 6.01, 8.01$ and $U=10$ used in previous sections, c.f., Figs.~\ref{fig:HDdistr}, \ref{fig:HHdistr}.}
\label{fig:gamma}
\end{figure}

The estimation of the recombination rate in the driven system is obtained by using the Floquet  Fermi golden rule developed in Refs.~\onlinecite{kitagawa2011,Bilitewski2015,ikeda2021}. The application to our example is derived in the Appendix~\ref{FloquetFGR}. The main conclusion is that the total recombination rate is a sum over decays in different channels 
\begin{equation}\label{Eq:ffgr}
\Gamma \approx \sum_{l \le l_0} g_{{\rm rc}}(l)^2 e^{-\xi} \sqrt{\frac{2\pi}{\Delta(l) \omega_0(l)}}\exp\left[-\frac{\Delta(l)}{\omega_0(l)} \ln\left(\frac{\Delta(l)}{e \xi \omega_0(l)}\right)\right],
\end{equation}
where the Mott gap size is reduced by the $l$-th Floquet sideband as $\Delta(l) \approx U-l\omega$ and the renormalized magnon frequency is given by $g_{\rm rc}(l)=\omega_0(l) \approx \frac{4t^2}{U-l\omega}\mathcal{J}_{l}(A)^2$. The sum in Eq.~\eqref{Eq:ffgr} needs to be restricted to $l \le l_0,$ where $l_0=\lfloor U/\omega\rfloor$ and $\lfloor\rfloor$ is the floor function. 
In such situation, the emission of energy of order $U$ in the recombination process is assisted by the driving via $l$-th Floquet sideband, such that only $\Delta(l)$ energy is emitted into spin excitations. For $l>l_0$, the excees energy $l\omega - U$ would have to be absorbed by the system, e.g.,  by absorptions of magnons, which is highly unlikely and we set this probability to zero.
%In such situation, the excess energy between the $l$-th Floquet sideband and $U$ can be absorbed by an emission of magnons. For $l>l_0$, the  excess magnons would need to be absorbed in a system which is a highly improbable event and we set its probability to zero. 
Furthermore, we assume that $\xi$ does not change with driving.

In Fig.~\ref{fig:gamma}, we present the decay rate for several driving frequencies versus the driving strength. In the weak driving regime, the recombination rate is supressed and we can consider Mott excitons as long lived states. At stronger excitations strenghts the recombination rate  exponentially increases and eventually excitons are not anymore well defined quantities. Comparison between Fig.~\ref{fig:gamma} and the previous analysis of increased binding shows that there exist a broad region of driving strengths $A$ where both excitons binding is increased and excitons still form long lived states. In particular, we find that for the driving frequency $\omega=6.01$, excitons are long lived states with the lifetime longer than $\tau>10^4$ over the whole range of driving amplitudes while their tendecy for binding is substantially increased, as shown in Fig.~\ref{fig:HDdistr}. This result proves that Floquet manipulation of holon-doublon pairing can lead to long-lived nonthermal state with enhanced pairing and should be considered as one of the most important messages of this paper.  While holon-holon bound states do not suffer from recombination, the opposite process of doublon-holon creation would plague the stability of holon-holon binding in the Floquet setup at the similar rate $\Gamma$.

\section{Photo-doped states in cold atoms experiments}
\label{coldatom}
While the existence of the doublon-holon excitons has been primarily discussed in the context of photo-doped solid-state setups~\cite{kishida2001,ono2004,lu2015,novelli2012,rincon2021}, cold-atom experiments could serve as an excellent testbed to study properties of photo-doped states. Recently, there has been tremendous progress in understanding half-filled~\cite{mazurenko2017} and doped Fermi-Hubbard systems with cold atom setups~\cite{bohrdt2021} exhibiting paradigmatic non-perturbative effects such as the pseudogap~\cite{brown2020}, spin strings~\cite{xu2023,chiu2019}, linear resistivity~\cite{brown2019}, etc. The application of the Floquet engineering to the chemically doped systems presented in this work can be straightforwardly extended to cold atom setups, either using electric field~(length) gauge~\cite{sandholzer19} or artificial gauge fields~\cite{aidelsburger18}. A similar superexchange manipulation can be obtained by directly modifying the hopping integral~\cite{gorg2018,Mentink2015}, $t_{ij}(\tau) = t_{ij}^0(1 + A\cos(\omega t + \phi_{ij}))$. In that case, the higher-order harmonics are suppressed, and we have checked that results remain qualitatively similar~(not shown).  A new avenue is the extension to photo-doped systems \cite{sandholzer19} and we propose simple protocols enabling a systematic study of differences between the chemically and photo-doped phase diagrams. We will consider two protocols to prepare the photo-doped state: the adiabatic state preparation from band to Mott insulator and (chirped) excitation across the Mott gap. 

\paragraph{Adiabatic state preparation}
In the first protocol, we start from a Mott-insulating state, where two sites have large positive and negative on-site potential, leading to fully occupied and empty states. Now, one can slowly reduce the on-site potential, allowing for the virtual dressing of the empty and full state with charge and spin fluctuations. The ramp velocity has to be slow enough so that the Landau-Zenner tunnelling across the Mott gap is suppressed~\cite{oka2005,oka2010,lenarcic2012}. With this respect, it might be convenient to first prepare the doublon-holon pair at high values of the Mott repulsion $U$ and then reduce it to the desired value. Such protocols can be easily extended to multiple holon-doublon pairs. This would allow us to study a complete phase diagram of photo-doped Mott insulators, including peculiar long-range phases such as the $\eta$-pairing superconductivity or the charge-density wave~\cite{murakami2022,Li2020}. 

\paragraph{Excitations accross the gap}
A standard approach to create holon-doublon pair is to either excite the system with a resonant excitation or apply an electric field comparable to the Mott gap. A naive implementation of the protocol would lead to a highly excited state with broad distribution of holes and doublon within the Hubbard band~\cite{oka2012} which is detrimental for the formation of Hubbard excitons with rather small binding energies and would require a substantial coupling with an external heat bath. Instead, we propose using a chirped pulse~(either electromagnetic or parameter modulation) with the base frequency slightly below the Mott gap, which is then slowly chirped across the Mott gap. It was shown in Ref.~\cite{werner2019a,werner2019,Li2020,werner2019} that such a protocol creates a quasi-equilibrium distribution with low effective temperatures and doublon-holon excitons should emerge as a stable structure.

Upon the successful preparation of the doublon-holon state through either of the prescribed protocols, one can initiate the application of the proposed Floquet engineering technique (Eq.~\eqref{Eq:Floquet}) to manipulate the exciton binding energy. 

As a consequence of state preparation and Floquet protocol, deviations from our ground state calculations could occur, most prominantly, finite temperature and finite charge density effects. Temperature effects can be effeciently mitigated using entropy cooling techniques~\cite{werner2019,nava2018} or chirped laser pulse~\cite{Li2020,werner2019b}. As shown in Ref.~\onlinecite{Mehio2023}, not only the lowest but also a few excited states at the edge of the upper Hubbard band have the nature of bound excitons, meaning one would still observe excitonic effect even if the effective temperature would be of a fraction of exchange coupling. Regarding the finite density of charges, we expect that densities up to a few percent should not significantly alter the results.

\section{Conclusions}
\label{diss}
In this work, we compared the pairing of chemically and photo-doped charge carriers mediated by spin and $\eta$ fluctuations in ladder and 2D systems. As the photo-doped state is metastable, we employed the Schrieffer-Wolf transformation to approximate the long-lived phase as a quasi-equilibrium state described by an extended t-J model. We establish that the binding is comparable in the two regimes by analysing binding energies and real-space correlators. In equilibrium and within the single-band Hubbard model, the superexchange $J$ and the hopping $t$ are connected and while the increase of the superexchange would enhance charge pairing it naturally comes with the increase of the hopping, leading to the tendency to delocalize charge carriers. We used Floquet engineering to decouple the two and the analysis yields an effective Floquet Hamiltonian, where both the superexchange $\mathfrak{J}$ and the hopping $\mathfrak{t}$ are modified by the driving frequency $\omega$ and the amplitude $A$. We demonstrate that a judicious selection of drive parameters can substantially enhance the strength of binding between exciton pairs both for ladders and 2D systems, although the increase in the latter is limited in a narrower driving range. The tendency to form bound pairs is most obvious in the analysis of the real-space correlators, which show a dramatic reduction in the relative distance between pairs. Finally, we provided an upper bound on the lifetime of the metastable phase using a Fermi golden rule argument and commented on how our theoretical predictions could be analyzed in cold-atom experiments.

Our estimation of the lifetime is based on Floquet Fermi golden rule, where we have
made several assumptions, like an usage of an approximate decay rate extracted from equilibrium decay rate~(Eq.~\ref{Eq:recombination}) and the time average approximation for rotating part of the recombination term, see Appendix~\ref{FloquetFGR}. It is an important future problem to relax these assumptions either by the usage of full Floquet Fermi-golden rule formalism or alternatively by a simulation of the full Hubbard model prepared with a holon-doublon pair and the inclusion of periodic driving. 
% An alternative approach is to generalize the Floquet Fermi golden rule~\cite{ikeda2021} approach for photo-doped Mott insulators, where the number of holons and doublons present an almost conserved quantity.

In the current work, we have focused on a dilute limit with a single pair; however, the formation of long-ranged orders will depend on the interaction between bound pairs. An exciting new perspective is to understand what is the effective interaction between holon-doublon pairs and what would be the symmetry of the $\eta$-paired condensate in the ladder and 2D systems. As the Floquet driving can change the pairing of bound pairs, it will also have an effect on the interaction between pairs and a systematic analysis of these processes is needed to understand how to stabilize non-thermal states with long-range orders.

\begin{acknowledgments}
We thank Y. Murakami, T. Kaneko, M. Bukov, and M. Eckstein for several useful discussions. We acknowledge the support by the projects J1-2463, N1-0318, MN-0016-106 and P1-0044 program of the Slovenian Research Agency, the QuantERA grant T-NiSQ by MVZI, QuantERA II JTC 2021, and ERC StG 2022 project DrumS, Grant Agreement 101077265. ED calculations were performed at the cluster `spinon' of JSI, Ljubljana. 
\end{acknowledgments}

\bibliography{Lib}

\appendix

\section{Effective Hamiltonian for the off-resonant regime}
\label{hamiloff}
Here, we show the various steps involved in arriving at the effective Hamiltonian $H_{\text{eff}}^{(F)}$ with the parameters shown in Eqs.~(\ref{r2exparam}-\ref{JzFloq}).
In the off-resonant regime, we first apply a unitary transformation by $\mathcal{U}(\tau) = \exp{(- i U \tau \sum_j n_{j, \uparrow} n_{j, \downarrow})}$. The resultant Hamiltonian in the rotated frame is given by $H^{{\rm rot}}(\tau)= \mathcal{U}^{\dagger}(\tau) H(\tau) \mathcal{U}(\tau) + i(\delta_{\tau} \mathcal{U}^{\dagger}(\tau)) \mathcal{U}(\tau)  $). $H^{{\rm rot}}(\tau)$ can be written as a sum of terms that preserve the number of DH pairs ($\Tilde{T}_0 (\tau)$), increase or decrease the DH pair number by one ($\Tilde{T}_1 (\tau)$, $\Tilde{T}_{-1} (\tau)$).
\begin{eqnarray}
    H^{{\rm rot}} (\tau)= - (\Tilde{T}_0 (\tau)+\Tilde{T}_1 (\tau)+\Tilde{T}_{-1} (\tau))
    \label{rotn1}
\end{eqnarray}
Introducing, $g_{ij\sigma}= (1- n_{i\bar{\sigma}}) c_{i \sigma}^{\dagger} c_{j \sigma} (1- n_{j\bar{\sigma}})+ n_{i\bar{\sigma}} c_{i \sigma}^{\dagger} c_{j \sigma} n_{j\bar{\sigma}} $ and $h_{i j \sigma}^{\dagger}= n_{i\bar{\sigma}} c_{i \sigma}^{\dagger} c_{j \sigma} (1- n_{j\bar{\sigma}}) $, the various terms in $H^{{\rm rot}}(\tau)$ can be written as
\begin{equation} 
\begin{split} 
 \Tilde{T}_0 (\tau) &= \sum_{\langle i,j\rangle,\sigma} t_{ij}(\tau) g_{ij\sigma}\\
 \Tilde{T}_1 (\tau) &= \sum_{\langle i,j\rangle,\sigma} t_{ij}(\tau) e^{iU\tau}h_{ij\sigma}^{\dagger} \\
 \Tilde{T}_{-1} (\tau)&= \sum_{\langle i,j\rangle,\sigma} t_{ij}^*(\tau) e^{-iU\tau} h_{ij\sigma}
    \end{split}
    \label{eq::rot1}
\end{equation}
Using $t_{ij}(\tau)= t_{ij} \sum_l A_{\mathbf{r}_{ij}}^{(l)} e^{i l \omega \tau}$ and Eq.~\eqref{eq::rot1} in Eq.~\eqref{rotn1},
\begin{align}
     H^{{\rm rot}} (\tau)&=- t_{ij} \sum_{\langle i,j\rangle,\sigma}   \sum_l \Big[A_{\mathbf{r}_{ij}}^{(l)} e^{i l \omega \tau} g_{ij\sigma} \label{eq::rot2} \\ 
     &+ A_{\mathbf{r}_{ij}}^{(l)} e^{i l \omega \tau} e^{iU\tau}h_{ij\sigma}^{\dagger} +  A_{\mathbf{r}_{ij}}^{(l)} e^{-i l \omega \tau} e^{-iU\tau} h_{ij\sigma} )\Big] \notag
\end{align}
Notice that $H^{{\rm rot}}(\tau)$ has two defining frequencies, U and $\omega$. To make $H^{{\rm rot}}(\tau)$ periodic in both U and $\omega$ we define a common frequency $\Omega_0$, such that $U=l_0 \Omega_0$ and $\omega=k_0 \Omega_0$, where $l_0$ and $k_0$ are co-prime numbers. To calculate the effective Floquet Hamiltonian we do a high-frequency expansion of $H^{{\rm rot}}(\tau)$ by expanding it with respect to different orders of the period $T=1/\Omega_0$. 
%as $\sum_{n=0}^{\infty} T^n H_{{\rm eff}}^{(n)}$. 
The first term in the expansion has the form \cite{Eckardt_2015},
\begin{eqnarray}\label{eq::mag1}
    H_{{\rm eff}}^{(0)} &=& \frac{1}{T} \int_0^T H^{{\rm rot}}(\tau_1) d\tau_1\\ \nonumber
    &=& H_{{\rm eff},1}^{(0)} + H_{{\rm eff},2}^{(0)}+H_{{\rm eff},2}^{(0)\dagger}
  % H_{{\rm eff}}^{(1)} &=& \frac{1}{2 i T^2} \int_0^T d\tau_1 \int_0^{\tau_1} [H^{{\rm rot}}(\tau_1),H^{{\rm rot}}(\tau_2)] d\tau_2
\end{eqnarray}

Now using Eq.~\eqref{eq::rot2} in Eq.~\eqref{eq::mag1}, the terms of the effective Floquet Hamiltonian at zeroth order are,
\begin{equation}
\begin{split}
    H_{{\rm eff},1}^{(0)} &= - \frac{1}{T} \int_{0}^T d\tau_1 \Tilde{T}_0 (\tau_1)  \\&=-\sum_{\langle i,j\rangle,\sigma} \frac{t_{ij}}{T} \int_0^T \sum_l A_{\mathbf{r}_{ij}}^{(l)}  e^{il\omega \tau_1} g_{ij\sigma} d\tau_1  \\ 
    &= - \sum_{\langle i,j\rangle,\sigma} t_{ij} A_{\mathbf{r}_{ij}}^{(0)} g_{ij\sigma}
    \end{split}
    \label{fzero}
    \end{equation}

\begin{align}
    H_{{\rm eff},2}^{(0)} &=- \frac{1}{T} \int_{0}^T d\tau_1 \Tilde{T}_1 (\tau)  \notag \\ &= -\sum_{\langle i,j\rangle,\sigma} \frac{t_{ij}}{T} \int_0^T \sum_l A_{\mathbf{r}_{ij}}^{(l)} e^{il\omega \tau_1} e^{i U \tau_1} h_{ij\sigma}^{\dagger} d\tau_1  \notag \\ 
    &=-\sum_{\langle i,j\rangle,\sigma} t_{ij} \sum_l A_{\mathbf{r}_{ij}}^{(l)} h_{ij\sigma}^{\dagger} \frac{1}{2 \pi} \int_0^{2 \pi} e^{i (lk_0+l_0) \Omega_0 \tau_1} d(\Omega_0 \tau_1)  \notag \\ 
    &=- \sum_{\langle i,j\rangle,\sigma} t_{ij} \sum_l A_{\mathbf{r}_{ij}}^{(l)} h_{ij\sigma}^{\dagger} \delta_{lk_0,-l_0}  \notag\\ &= 0  \label{recomzero}
\end{align}
Since $l_0$ and $k_0$ are co-prime numbers, $l_0 \neq l k_0$, and the above integral is zero. Similarly for the complex conjugate part $H_{{\rm eff},2}^{(0)\dagger}=0$. Hence,  at the zeroth order in the high-frequency expansion the effective Hamiltonian is,
\begin{eqnarray}
    H_{{\rm eff}}^{(0)} = \sum_{\langle i,j\rangle,\sigma} t_{ij} A_{\mathbf{r}_{ij}}^{(0)} g_{ij\sigma}
\end{eqnarray}

The next order term is given by
\begin{equation}
\begin{split}
    H_{{\rm eff}}^{(1)} &= \frac{1}{T^2} \int_0^T d\tau_1 \int_0^{T} d\tau_2 \sum_{m \neq 0} \frac{e^{- i m \Omega_0 (\tau_1 - \tau_2)}}{m \Omega_0} H^{{\rm rot}}(\tau_1)~H^{{\rm rot}}(\tau_2)\\ &=  H_{{\rm eff},1}^{(1)}+  H_{{\rm eff},2}^{(1)}+ H_{{\rm eff},2}^{(1) \dagger}+H_{{\rm eff},3}^{(1)}+ H_{{\rm eff},3}^{(1) \dagger}
    \end{split},
\end{equation}
The first term is given by,
\onecolumngrid
\vspace{0.5\columnsep}
\begin{equation}
\begin{split}
    H_{{\rm eff},1}^{(1)} &= \frac{1}{T^2} \int_0^T d\tau_1 \int_{0}^{\tau_1} d\tau_2~ \sum_{m \neq 0} \frac{e^{- i m \Omega_0 (\tau_1 - \tau_2)}}{m \Omega_0} \left[\Tilde{T}_0 (\tau_1),\Tilde{T}_0 (\tau_2)\right]\\  &=\frac{1}{T^2} \int_0^T d\tau_1 \int_0^{\tau_1} d\tau_2~\sum_{m \neq 0} \frac{e^{- i m \Omega_0 (\tau_1 - \tau_2)}}{m \Omega_0} \left[ \sum_{\langle i,j \rangle,\sigma,l} t_{ij} A_{\mathbf{r}_{ij}}^{(l)} e^{il \omega \tau_1} g_{ij\sigma}  , \sum_{\langle i^{\prime},j^{\prime} \rangle,\sigma^{\prime},l^{\prime}} t_{i^{\prime}j^{\prime}} A_{\mathbf{r}_{i^{\prime}j^{\prime}}}^{(l^{\prime})} e^{i l^{\prime} \omega \tau_2} g_{i^{\prime}j^{\prime}\sigma^{\prime}}\right]  = 0\\
\end{split}
    \end{equation}
since all the commutators vanish. The second term is given by,
\begin{equation}
    \begin{split}
    H_{{\rm eff},2}^{(1)} &= \frac{1}{T^2} \int_0^T d\tau_1 \int_{0}^{T} d\tau_2  ~ \sum_{m \neq 0} \frac{e^{- i m \Omega_0 (\tau_1 - \tau_2)}}{m \Omega_0}  \Tilde{T}_0 (\tau_1) \Tilde{T}_1 (\tau_2) \\ &= \sum_{m \neq 0} \frac{1}{ m \Omega_0 T^2} \int_0^T d\tau_1 \int_0^{T} d\tau_2  e^{- i m \Omega_0 (\tau_1 - \tau_2)}   \sum_{ \langle i,j \rangle,\sigma,l} t_{ij} A_{\mathbf{r}_{ij}}^{(l)} e^{il k_0 \Omega_0 \tau_1} g_{ij\sigma}  \sum_{ \langle i^{\prime},j^{\prime} \rangle,\sigma^{\prime}, l^{\prime}} t_{i^{\prime}j^{\prime}}  A_{\mathbf{r}_{i^{\prime} j^{\prime}}}^{(l^{\prime})} e^{-i (l^{\prime} k_0 \Omega_0+ l_0 \Omega_0) \tau_2} h_{i^{\prime}j^{\prime}\sigma^{\prime}} \\ =&  \sum_{m \neq 0} \sum_{ \langle i,j \rangle,\sigma,l} ~\sum_{ \langle i^{\prime},j^{\prime} \rangle,\sigma^{\prime}, l^{\prime}} \frac{t_{ij}~ t_{i^{\prime}j^{\prime}}~A_{\mathbf{r}_{ij}}^{(l)} ~ A_{\mathbf{r}_{i^{\prime} j^{\prime}}}^{(l^{\prime})}}{m \Omega_0} \frac{1}{2 \pi}\int_0^{2 \pi} d(\Omega_0 \tau_1) ~e^{i (l k_0-m) \Omega_0 \tau_1} \frac{1}{2 \pi}\int_0^{2 \pi} d (\Omega_0 \tau_2) ~e^{i (-l^{\prime} k_0 - l_0+m ) \Omega_0 \tau_2} ~g_{ij\sigma} ~h_{i^{\prime}j^{\prime}\sigma^{\prime}} \\ =& \sum_{m \neq 0} \sum_{ \langle i,j \rangle,\sigma,l} ~\sum_{ \langle i^{\prime},j^{\prime} \rangle,\sigma^{\prime}, l^{\prime}} \frac{t_{ij}~ t_{i^{\prime}j^{\prime}}~A_{\mathbf{r}_{ij}}^{(l)} ~ A_{\mathbf{r}_{i^{\prime} j^{\prime}}}^{(l^{\prime})}}{m \Omega_0} ~ \delta_{l k_0,m} ~\delta_{l^{\prime} k_0+l_0,m}~g_{ij\sigma} ~h_{i^{\prime}j^{\prime}\sigma^{\prime}}
    \end{split}
\end{equation}
This integral is non-zero only when $m=l k_0$, and $m= l^{\prime} k_0 + l_0$. This gives, $l k_0 = l^{\prime} k_0 + l_0$ or $(l-l^{\prime}) k_0 = l_0$, which is never satisfied since $l_0$ and $k_0$ are co-primes. As a result, $H_{{\rm eff},2}^{(1)}=0$ always. Similarly,
\begin{equation} H_{{\rm eff},2}^{(1) \dagger} = \frac{1}{T^2} \int_0^T d\tau_1 \int_{0}^{T} d\tau_2  ~\frac{e^{- i m \Omega_0 (\tau_1 - \tau_2)}}{m \Omega_0}  \Tilde{T}_0 (\tau_1) \Tilde{T}_{-1} (\tau_2)=0
\end{equation}
The only non-zero integral at this order is,
\begin{equation}
\begin{split}
H_{{\rm eff},3}^{(1)}&= \frac{1}{T^2} \int_0^T d\tau_1 \int_{0}^{T} d\tau_2~\sum_{m \neq 0} \frac{e^{- i m \Omega_0 (\tau_1 - \tau_2)}}{m \Omega_0}  \Tilde{T}_1 (\tau_1)\Tilde{T}_{-1} (\tau_2)  \\ &=\sum_{m \neq 0}\frac{1}{m \Omega_0 T^2} \int_0^T d\tau_1 \int_0^{T}  d\tau_2~e^{- i m \Omega_0 (\tau_1 - \tau_2)} \sum_{\langle i,j \rangle,\sigma,l} t_{ij} A_{\mathbf{r}_{ij}}^{(l)}  e^{i(l k_0+l_0) \Omega_0 \tau_1} h_{ij\sigma}^{\dagger}  \sum_{\langle i^{\prime},j^{\prime} \rangle,\sigma^{\prime},l^{\prime}} t_{i^{\prime} j^{\prime}}A_{\mathbf{r}_{i^{\prime}j^{\prime}}}^{(l^{\prime})} e^{-i (l^{\prime} k_0+l_0) \Omega_0 \tau_2} h_{i^{\prime}j^{\prime}\sigma^{\prime}} \\ &= \sum_{m \neq 0} \sum_{ \langle i,j \rangle,\sigma,l} ~\sum_{ \langle i^{\prime},j^{\prime} \rangle,\sigma^{\prime}, l^{\prime}} \frac{t_{ij}~ t_{i^{\prime}j^{\prime}}~A_{\mathbf{r}_{ij}}^{(l)} ~ A_{\mathbf{r}_{i^{\prime} j^{\prime}}}^{(l^{\prime})}}{m \Omega_0} \frac{1}{2 \pi}\int_0^{2 \pi} d(\Omega_0 \tau_1) ~e^{i (l k_0+l_0-m) \Omega_0 \tau_1} \frac{1}{2 \pi}\int_0^{2 \pi} d (\Omega_0 \tau_2) ~e^{i (-l^{\prime} k_0 - l_0+m ) \Omega_0 \tau_2} ~h_{ij\sigma}^{\dagger} ~h_{i^{\prime}j^{\prime}\sigma^{\prime}} \\ &= \sum_{m \neq 0} \sum_{ \langle i,j \rangle,\sigma,l} ~\sum_{ \langle i^{\prime},j^{\prime} \rangle,\sigma^{\prime}, l^{\prime}} \frac{t_{ij}~ t_{i^{\prime}j^{\prime}}~A_{\mathbf{r}_{ij}}^{(l)} ~ A_{\mathbf{r}_{i^{\prime} j^{\prime}}}^{(l^{\prime})}}{m \Omega_0} ~ \delta_{l k_0+l_0,m} ~\delta_{l^{\prime} k_0+l_0,m} ~h_{ij\sigma}^{\dagger} ~h_{i^{\prime}j^{\prime}\sigma^{\prime}}\\ &=\sum_{ \langle i,j \rangle,\sigma,l} ~\sum_{ \langle i^{\prime},j^{\prime} \rangle,\sigma^{\prime}} \frac{t_{ij}~ t_{i^{\prime}j^{\prime}}~A_{\mathbf{r}_{ij}}^{(l)} ~ A_{\mathbf{r}_{i^{\prime} j^{\prime}}}^{(l)}}{(l k_0 + l_0) \Omega_0} ~h_{ij\sigma}^{\dagger} ~h_{i^{\prime}j^{\prime}\sigma^{\prime}} \\ &=\sum_{ \langle i,j \rangle,\sigma,l} ~\sum_{ \langle i^{\prime},j^{\prime} \rangle,\sigma^{\prime}} \frac{t_{ij}~ t_{i^{\prime}j^{\prime}}~A_{\mathbf{r}_{ij}}^{(l)} ~ A_{\mathbf{r}_{i^{\prime} j^{\prime}}}^{(l)}}{(U + l \omega)} ~h_{ij\sigma}^{\dagger} ~h_{i^{\prime}j^{\prime}\sigma^{\prime}}\\
%&= \sum_{l} \frac{1}{(U+ l \omega)}  \left[\sum_{\langle i,j \rangle,\sigma} t_{ij} A_{\mathbf{r}_{ij}}^{(l)} h_{ij\sigma}^{\dagger}, \sum_{\langle i^{\prime},j^{\prime} \rangle,\sigma^{\prime}} t_{i^{\prime}j^{\prime}} A_{\mathbf{r}_{i^{\prime}j^{\prime}}}^{(l)*} h_{i^{\prime}j^{\prime}\sigma^{\prime}}  \right] \\ &= (-1)\sum_{\langle i,j \rangle,\sigma} \sum_{l} \frac{4 t_{ij}^2}{U - l \omega} A_{\mathbf{r}_{ij}}^{-l} A_{-\mathbf{r}_{ij}}^{-l} \left( \mathbf{\eta}_i^x\mathbf{\eta}_j^{x} + \mathbf{\eta}_i^y\mathbf{\eta}_j^{y}  \right) + \sum_{\langle i,j \rangle,\sigma} \sum_l \frac{4 t_{ij}^2}{U - l \omega} \mid A_{\mathbf{r}_{ij}}^{(-l)} \mid ^2 \left( \mathbf{S}_i \mathbf{S}_j - \mathbf{\eta}_i^z\mathbf{\eta}_j^{z}  \right)
\end{split}
\label{f1st}
\end{equation}
Similarly the complex conjugate gives,
\begin{equation}
    \begin{split}
        H_{{\rm eff},3}^{(1) \dagger}&= \frac{1}{T^2} \int_0^T d\tau_1 \int_{0}^{T} d\tau_2~\sum_{m \neq 0} \frac{e^{- i m \Omega_0 (\tau_1 - \tau_2)}}{m \Omega_0}  \Tilde{T}_{-1} (\tau_1)\Tilde{T}_{1} (\tau_2)\\ &= - \sum_{ \langle i,j \rangle,\sigma,l} ~\sum_{ \langle i^{\prime},j^{\prime} \rangle,\sigma^{\prime}} \frac{t_{ij}~ t_{i^{\prime}j^{\prime}}~A_{\mathbf{r}_{ij}}^{(l)} ~ A_{\mathbf{r}_{i^{\prime} j^{\prime}}}^{(l)}}{(U + l \omega)} ~h_{ij\sigma} ~h_{i^{\prime}j^{\prime}\sigma^{\prime}}^{\dagger}
    \end{split}
    \label{f1sta}
\end{equation}
Combining Eq. ~\eqref{f1st} and Eq. ~\eqref{f1sta} and changing $l \rightarrow -l$ the Hamiltonian at this order is,
\begin{equation}
    \begin{split}
        H_{{\rm eff}}^{(1)}&= \sum_{ \langle i,j \rangle,\sigma,l} ~\sum_{ \langle i^{\prime},j^{\prime} \rangle,\sigma^{\prime}} \frac{t_{ij}~ t_{i^{\prime}j^{\prime}}~A_{\mathbf{r}_{ij}}^{(-l)} ~ A_{\mathbf{r}_{i^{\prime} j^{\prime}}}^{(-l)}}{(U - l \omega)} \left[h_{ij\sigma}^{\dagger},h_{i^{\prime}j^{\prime}\sigma^{\prime}}\right]
        \\ &= -\sum_{\langle i,j \rangle,\sigma} \sum_{l} \frac{4 t_{ij}^2}{U - l \omega} A_{\mathbf{r}_{ij}}^{-l} A_{-\mathbf{r}_{ij}}^{-l} \left( \mathbf{\eta}_i^x\mathbf{\eta}_j^{x} + \mathbf{\eta}_i^y\mathbf{\eta}_j^{y}  \right) + \sum_{\langle i,j \rangle,\sigma} \sum_l \frac{4 t_{ij}^2}{U - l \omega} \mid A_{\mathbf{r}_{ij}}^{(-l)} \mid ^2 \left( \mathbf{S}_i \cdot\mathbf{S}_j - \mathbf{\eta}_i^z\mathbf{\eta}_j^{z}  \right)
    \end{split}
\end{equation}
The last line follows from the explicit computation of the commutators \cite{murakami2023suppression,Bukov2016}.
So the effective Floquet Hamiltonian from Eq.~\eqref{fzero} and Eq.~\eqref{f1st} is,
\begin{equation}
\begin{split}
    H_{\text{eff}}^{(F)} &= \sum_{\langle i,j\rangle,\sigma} t_{ij} A_{\mathbf{r}_{ij}}^{(0)} g_{ij\sigma} - \sum_{\langle i,j\rangle,\sigma} \sum_{l} \frac{4 t_{ij}^2}{U - l \omega} A_{\mathbf{r}_{ij}}^{-l} A_{-\mathbf{r}_{ij}}^{-l} \left( \mathbf{\eta}_i^x\mathbf{\eta}_j^{x} + \mathbf{\eta}_i^y\mathbf{\eta}_j^{y}  \right) + \sum_{\langle i,j \rangle,\sigma} \sum_l \frac{4 t_{ij}^2}{U - l \omega} \mid A_{\mathbf{r}_{ij}}^{(-l)} \mid ^2 \left( \mathbf{S}_i\cdot\mathbf{S}_j -\mathbf{\eta}_i^z\mathbf{\eta}_j^{z}  \right)
    \end{split}
\end{equation}
\vspace{\columnsep}
\twocolumngrid

\section{Computation of ground state energy for the $2\times2$ cluster in the limit $\mathfrak{J}_{\perp}\gg \mathfrak{J}_{\parallel}$.}
\label{appcluster}
In this section, we present the computation of the ground state energy of systems consisting of one HH and DH pair, respectively. We will consider $2\times 2$ clusters, which allow for analytical results, in generic situation where the exchange parameters are distinguished by the Floquet drive as in Eqs.~(\ref{r2exparam}-\ref{JzFloq}). We will work in the limit $\mathfrak{J}_{\perp}\gg \mathfrak{J}_{\parallel}$ and neglect the hopping terms to get analytical results. Under these assumptions, the ground state manifold for one HH pair contains two basis states connected by the limited Hamiltonian, $|\frac{ 0 \uparrow }{ 0 \downarrow} \rangle$ and $|\frac{ 0 \downarrow }{ 0 \uparrow} \rangle$,
where the top row represents the spin configuration in the upper chain and the bottom in the lower chain. The Hamiltonian for the system is given by:\\
\begin{eqnarray*}
    \left(
\begin{array}{cc}
 -\frac{\mathfrak{J}_{\perp,S}}{2} & \frac{\mathfrak{J}_{\perp,S}}{2} \\
 \frac{\mathfrak{J}_{\perp,S}}{2} & -\frac{\mathfrak{J}_{\perp,S}}{2} \\
\end{array}
\right)
\end{eqnarray*}
From this, one can find the ground state energy of this manifold to be $\tilde{E}_{0}^{HH}=-\mathfrak{J}_{\perp,S}$ if $\mathfrak{J}_{\perp,S}>0$ and zero otherwise.\\\\
For the case of one DH pair, we work in the manifold with four states, 
$| \frac{0 \uparrow}{\updownarrow \downarrow} \rangle$, $\mid \frac{0 \downarrow}{\updownarrow \uparrow} \rangle$, $\mid \frac{\updownarrow \uparrow}{ 0 \downarrow }\rangle$ and $\mid \frac{\updownarrow \downarrow}{ 0 \uparrow} \rangle$. The Hamiltonian for this scenario is given by:
\begin{eqnarray*}
    \left(
\begin{array}{cccc}
 \frac{\mathfrak{J}_{\perp,\eta}^z}{2}-\frac{\mathfrak{J}_{\perp,S}}{2} & \frac{\mathfrak{J}_{\perp,S}}{2} & \frac{\mathfrak{J}_{\perp,\eta}^{xy}}{2} & 0 \\
 \frac{\mathfrak{J}_{\perp,S}}{2} & \frac{\mathfrak{J}_{\perp,\eta}^z}{2}-\frac{\mathfrak{J}_{\perp,S}}{2} & 0 & \frac{\mathfrak{J}_{\perp,\eta}^{xy}}{2} \\
 \frac{\mathfrak{J}_{\perp,\eta}^{xy}}{2} & 0 & \frac{\mathfrak{J}_{\perp,\eta}^z}{2}-\frac{\mathfrak{J}_{\perp,S}}{2} & \frac{\mathfrak{J}_{\perp,S}}{2} \\
 0 & \frac{\mathfrak{J}_{\perp,\eta}^{xy}}{2} & \frac{\mathfrak{J}_{\perp,S}}{2} & \frac{\mathfrak{J}_{\perp,\eta}^z}{2}-\frac{\mathfrak{J}_{\perp,S}}{2} \\
\end{array}
\right),
\end{eqnarray*}
where the diagonal elements are zero as $\mathfrak{J}_{\perp,\eta}^z=\mathfrak{J}_{\perp,S}.$ 
The eigenenergies are $\{-(\mathfrak{J}_{\perp,\eta}^{xy}+\mathfrak{J}_{\perp,S})/2,(\mathfrak{J}_{\perp,\eta}^{xy}-\mathfrak{J}_{\perp,S})/2,(-\mathfrak{J}_{\perp,\eta}^{xy}+\mathfrak{J}_{\perp,S})/2,(\mathfrak{J}_{\perp,\eta}^{xy}+\mathfrak{J}_{\perp,S})/2\}$. In equilibrium and for weak drivings, the state with energy $-(\mathfrak{J}_{\perp,\eta}^{xy}+\mathfrak{J}_{\perp,S})/2$ has the lowest energy and the corresponding eigenstate is given by $\ket{\phi}= \left(\mid \frac{\updownarrow \uparrow}{ 0 \downarrow }\rangle- |\frac{0 \uparrow}{\updownarrow \downarrow} \rangle + \mid \frac{0 \downarrow}{\updownarrow \uparrow} \rangle - \mid \frac{\updownarrow \downarrow}{ 0 \uparrow}\rangle \right)/2.$ At higher driving amplitudes $A$, other states can become the ground state and their ground state $\tilde{E}^{DH}_{0}(A)$ dependence is depicted in Fig.~\ref{fig:fig10}(d).
\section{Dominant recombination term in Floquet driven setup}
\label{apprecomb}
To establish an approximate recombination rate in the Floquet driven setup with the help of the static expression, Eq.~\eqref{Eq:recombination},  we perform a different time dependent canonical transformation which, similar to the static case, transforms out recombination term in the lowest order but retains it in the next order.

Starting from the original Hamiltonian
\begin{equation}
    \begin{split}
        H (\tau) &= - \sum_{ \ave{i,j} , \sigma} t_{ij} (\tau)  ( c^{\dagger}_{i \sigma} c_{j \sigma} + c^{\dagger}_{j \sigma } c_{i \sigma } ) + U \sum_{j} n_{ j \uparrow } n_{ j \downarrow} \\
        &\equiv - \lambda T (\tau) + U D, \quad \lambda=1,
    \end{split}
\end{equation}
where we again split the hopping term into contributions that either retain, increase or decrease the number of DH pairs,
$T(\tau) = \sum_l (T_{0,l} + T_{1,l}+ T_{-1,l}) e^{ i l \omega \tau}$, introduce a bookkeeping parameter $\lambda$, and define $D= \sum_{j} n_{ j \uparrow } n_{ j \downarrow}$ as the double occupancy operator.

Following Ref.~\onlinecite{Kitamura2017} we introduce a time-dependent canonical transformation $e^{i S(\tau)}$, and obtain the transformed Hamiltonian $H_{{\rm r}}$ as,
\begin{eqnarray}
   H_{{\rm r}} (\tau) +  U D = e^{i S(\tau) } \left[ H(\tau) - i \partial_{\tau} \right] e^{ - i S(\tau) }
   \label{eq::sce1}
\end{eqnarray}
$S (\tau)$ is chosen to eliminate the terms off-diagonal in $D$. To determine $H_{{\rm r}} (\tau)$ and  $S (\tau)$ order by order we expand them in powers of $\lambda$ as, $H_{{\rm r}} (\tau) = \sum_{n=1}^{\infty} \lambda^n H_{{\rm r}}^{(n)}$ and $S (\tau) = \sum_{n=1}^{\infty} \lambda^n S^{(n)}$. We can re-write Eq.~\eqref{eq::sce1} as
\begin{equation}
\begin{split}
    H_{{\rm r}} (\tau) &+ U \left[D, i S (\tau) \right] + i \partial_{\tau} S (\tau) \\ &= - \lambda T (\tau) - \sum_{n=0}^{\infty} \frac{B_n}{n!} {\rm ad}^{n}_{iS}\left[(-1)^{(n)} \lambda T(\tau) + H_{{\rm r}}(\tau) \right].
    \end{split}
    \label{eq::sce2}
\end{equation}
%We truncate $S (\tau) = \lambda S^{(1)}$, since 
where $\rm ad_A B = [A,B]$ and $B_n$ is the Bernoulli number.
Our aim is to transform out the recombination at order $\mathcal{O}(\lambda)$, but keep it in the second order $\mathcal{O}(\lambda^2)$ so as to compare the Floquet expression to the static one. From Eq.~\eqref{eq::sce2},
\begin{equation}
    \begin{split}
        H_{{\rm r}}^{(1)} &= - T - U \left[ D , i S^{(1)} \right] - \partial_{\tau} S^{(1)},
    \end{split}
\end{equation}
we find that $i S^{(1)} = - \sum_l (\frac{T_{1,l}}{U+l\omega} - \frac{T_{-1,l}}{U-l\omega}) e^{i l \omega \tau}$ gives $H_{{\rm r}}^{(1)}= - \sum_l T_{0,l} e^{ i l\omega \tau}$, i.e., it eliminates the recombination in the lowest order.

From Eq.~\eqref{eq::sce2} and truncating $S^{(k>1)}=0$ we get
\begin{equation}
    \begin{split}
        H_{{\rm r}}^{(2)} &= - \frac{1}{2} \left[ iS^{(1)}, T-H_{{\rm r}}^{(1)}\right]\\
        &= - \frac{1}{2} \left[ iS^{(1)} , \sum_l (T_{-1,l} + 2 T_{0,l} + T_{1,l}) e^{il\omega\tau} \right]
    \end{split}
\end{equation}
The relevant term in the commutator causing the recombination is given by
\begin{equation}
    \begin{split}
        &-\frac{1}{2} \left[ \sum_l \frac{T_{-1,l}}{U-l\omega} e^{i l\omega\tau}, 2 \sum_{l^{\prime}} T_{0,l^{\prime}} e^{i l^{\prime} \omega\tau} 
  \right] \\
  &= -\sum_{l,l^{\prime}} \frac{e^{i(l+l^{\prime})\omega\tau}}{U-l\omega} \left[ T_{-1,l}, T_{0,l^{\prime}} \right]
    \end{split}
\end{equation}
The effect of this term in the lowest order of the high-frequency expansion is given by its time average over one time period,
\begin{equation}
    \begin{split}
       H_{{\rm rc}}^{(F)} &=-\frac{1}{T} \int_0^T d\tau \sum_{l,l^{\prime}} \frac{e^{i(l+l^{\prime})\omega \tau}}{U-l\omega} \left[ T_{-1,l}, T_{0,l^{\prime}} \right] \\
        &= \sum_{l} \frac{1}{U-l \omega} \left[ T_{-1,l}, T_{0,-l} \right]
    \end{split}
    \label{eq:recom1}
\end{equation}
We can explicitly write down $T_{0,l},T_{-1,l},T_{1,l}$  using $g_{ij\sigma}$, $h_{ij\sigma}$ defined in Appendix~\ref{hamiloff} as
\begin{equation}
    \begin{split}
        T_{0,l} &= \sum_{\langle i,j\rangle, \sigma}  t_{ij} \mathcal{J}_{l} (\mathbf{r}_{ij}\cdot\mathbf{A}) g_{ij \sigma} \\
        T_{1,l} &= \sum_{\langle i, j\rangle, \sigma}  t_{ij} \mathcal{J}_{l} (\mathbf{r}_{ij}
        \cdot\mathbf{A}) h_{ij \sigma}^{\dagger} \\
        T_{-1,l} &= \sum_{\langle i, j\rangle, \sigma}  t_{ij} \mathcal{J}_{l} (\mathbf{r}_{ji}\cdot\mathbf{A}) h_{ij \sigma} 
    \end{split}
    \label{eq::hopeq}
\end{equation}
Then by using Eq.~\eqref{eq:recom1}, we find that the recombination term is given by,
\begin{equation}
\begin{split}
    H_{{\rm rc}}^{(F)} &= -\sum_{ijk,ss'} \sum_{l} \frac{t_{ij} t_{jk}}{U- l \omega} \mathcal{J}_{l} (\mathbf{r}_{ji}\cdot\mathbf{A}) \mathcal{J}_{-l} (\mathbf{r}_{jk}\cdot\mathbf{A}) \left[ h_{ijs}, g_{jks'} \right] \\
    &=\frac{1}{2} \sum_l \sum_{(ijk)ss'}
\frac{t_{ij} t_{jk} 
\mathcal{J}_{l}( \boldsymbol{r}_{ji} \cdot \boldsymbol{A} ) 
\mathcal{J}_{-l}(\boldsymbol{r}_{jk} \cdot \boldsymbol{A} )}{U-l \omega} \\ 
&\hspace{3cm}\times(1-n_{i,\bar{s}}) c_{is}^{\dagger} n_{ks'}c_{k\bar{s}'} \, \vec{\sigma}_{s \bar{s}'}\cdot \boldsymbol{S}_j 
\end{split}
\end{equation}
which is Eq.~\eqref{eq::HrcF} in Sec.~\ref{sec:recom}.

\section{Computation of recombination rate in the presence of Floquet drive}\label{FloquetFGR}
In this appendix, we derive the estimate for the recombination rate in the periodically driven system based on Floquet Fermi golden rule~\cite{kitagawa2011,Bilitewski2015,ikeda2021}. We follow the spirit of Ref.~\onlinecite{Bilitewski2015} , but in our case both, charge conserving and recombination/creation terms, are time dependent.

Since we apply a time periodic drive we have,
\begin{equation}
    H(t+T) = H(t),
\end{equation}
where $T$ is the period of oscillation and the solution is given by time equivalent of the Bloch state $\Psi_{\alpha}(t)= e^{-\frac{i \epsilon_{\alpha} t}{\bar{h}}} \Phi_{\alpha}(\tau)$, where a Floquet mode has periodic property $\Phi_{\alpha}(t+T)=\Phi_{\alpha}(t)$ and $\epsilon_{\alpha}$ are the eigen-energies defined up to multiples of $\hbar \omega,$ where $\omega=2\pi/T.$ We can expand these states as 
\begin{equation}
\Phi_{\alpha}(t)=\sum_m e^{i m \omega t} \phi_{\alpha,m}(t),
\end{equation}
where $\phi_{\alpha,m}$ is the $m$-th Fourier mode. Floquet modes also forms a complete and orthonormal basis set 
\begin{equation}
    \langle\langle \Phi^n_{\alpha}|\Phi^m_{\beta} \rangle\rangle=\frac{1}{T}\int_0^T dt \langle \Phi^n_{\alpha}|\Phi^m_{\beta} \rangle=\delta_{\alpha\beta}\delta_{mn}
\end{equation}

% Then if we consider $\psi(\tau)$ such that, $(H(\tau)- i \frac{i \delta}{\delta~\tau})=0$ we can say, $\psi_{\alpha}(\tau)= e^{\frac{i \epsilon_{\alpha} t}{\bar{h}}} \phi_{\alpha}(\tau)$, where $\phi_{\alpha}(\tau+T)=\phi_{\alpha}(\tau)$ are "Bloch states" and $\phi_{\alpha}(\tau)= e^{i m \omega \tau} \phi_{\alpha}$.\\

The essential idea behind the estimation of doublon-holon recombination time in Ref.~\onlinecite{lenarcic13} is to use the Fermi golden rule, where we separate the  Hamiltonian into doublon-holon conserving part $H_0(t)$ and recombination term $V(t)$. Now, we can separate the evolution into the charge conserving part $U_0(t,t_0)$, governed by $H_0(t)$. The full evolution is given by $U(t,t_0)=U_0(t,t_0)U_I(t,t_0),$ where $U_I(t,t_0)$
is the propagator in the interaction picture expressed as 
$U_I(t,t_0)=1-\frac{i}{\hbar}\int_{t_0}^t dt' V_I(t')U_I(t',t_0)$ and $V_I(t)=U_0(t_0,t)V(t) U_0(t,t_0)$.
The latter can be simplified by taking the first order expansion in the Dyson series $U_I(t,t_0)=1-\frac{i}{\hbar}\int_{t_0}^t dt' V_I(t')$.

We are intested in estimating the probability for transition between the initial state with holon and doublon pair $\Psi_i(t_0=0)$ to the final state $\Psi_f(t_f=\tau)$ where the pair has recombined due recombination term $V(t)$. The relevant amplitude for the transition is given by 
 \begin{eqnarray}
     \rho_{i \rightarrow f} (\tau) &=& -i \int_0^{\tau} d\tau'  \langle \Psi_f(0) | V_I(\tau') | \Psi_i (0) \rangle\\ \nonumber
   &=&  -i \int_0^{\tau} d\tau^{\prime}  \langle \Psi_f (0)  | \mathcal{U}_0(0,\tau^{\prime}) V(\tau^{\prime}) \mathcal{U}_0(\tau^{\prime},0)|\Psi_i (0) \rangle \\ \nonumber
   &=&  -i \int_0^{\tau} d\tau^{\prime} e^{- i (\epsilon_{i} - \epsilon_{f}) \tau^{\prime}}  \langle \Phi_f (\tau^{\prime})  |  V(\tau^{\prime}) | \Phi_i (\tau^{\prime}) \rangle \\ \nonumber
    &=&  -i \sum_{\ell, n, m} \int_0^{\tau} d\tau^{\prime} e^{- i (\epsilon_{i} - \epsilon_{f} - (n-m+\ell)\omega) \tau^{\prime}} \\ \nonumber  &&\langle\phi_{f,m}  |  V^{\ell} | \phi_{i,n} \rangle \\ \nonumber
    &=& \sum_{\ell, n, m} \frac{e^{- i (\epsilon_{i} - \epsilon_{f} - (n-m+\ell)\omega) \tau^{\prime}} -1}{\epsilon_{i} - \epsilon_{f} - (n-m+\ell)\omega}~ V_{m,n,\ell}^{fi}
 \end{eqnarray}

 where we have defined the  matrix element of the $\ell$-th Floquet mode of the recombination term $V(\tau)=\sum_{\ell} e^{i \omega \ell \tau} V^{\ell}$ as $V_{m,n,\ell}^{fi}=\langle\phi_{f,m}  |  V^{\ell} | \phi_{i,n} \rangle.$  However, in the following we will consider only the average $\ell=0$ component.
 
The transition probability is given by 

 \begin{eqnarray}
     \Gamma_{i \rightarrow f }&=& \lim_{\tau \rightarrow \infty} \frac{|\rho_{i \rightarrow f}(\tau)|^2}{\tau} \hspace{1.7in} ({\rm C}5)\nonumber \\ \nonumber
     &=& 2 \pi \sum_{n, m, p}  \delta (\epsilon_i - \epsilon_f - p \omega)  V^{fi}_{n,n+p,0} V^{if}_{m+p,m,0} \\ \nonumber 
     &=& 2 \pi \sum_{n,m,p} \delta (\epsilon_i - \epsilon_f - p \omega) \frac{(t ~\mathcal{J}_{p} )^2 (t~ \mathcal{J}_{-p} )^2}{(U - p \omega)^2}  \tilde V^{fi}_{n,n+p} \tilde V^{if}_{m+p,m}\\
     &=&2\pi
     \sum_{p} \delta (\epsilon_i^0 - \epsilon_f^0 - p \omega) \frac{(t ~\mathcal{J}_{p} )^2 (t~ \mathcal{J}_{-p} )^2}{(U - p \omega)^2} |\langle \Phi_f^p|\tilde V|\Phi_i^0 \rangle|^2\nonumber
 \end{eqnarray} 

where we have separated out the time-independent operatorial part $\tilde V=\sum_{ijk} (1-n_{i,\bar{s}}) c_{is} n_{ks'}c_{k\bar{s}'} \, \vec{\sigma}_{s \bar{s}'}\cdot \boldsymbol{S}_j$ of the recombination term. Here, $\Phi^p_{i,f}(t)=e^{i p\omega  t}\Phi^0_{i,f}(t)$ are shifted Floquet modes and eigenenergies $\epsilon_{\alpha}^0$ fulfill $-\hbar \omega/2<\epsilon_{\alpha}^0<\hbar \omega/2$. The final expression allow us to use estimation based on equilibrium Fermi golden rule of Ref.~\onlinecite{lenarcic13} generalized for each Floquet channel separately.

\section{Role of $\eta$-pairing term in exciton binding}\label{eta}
In this appendix, we compare and contrast the scenario of binding between Hubbard exciton pairs in the presence and absence of the $\eta$-pairing term, $H_{\eta}^{\parallel (\perp)}$.

\begin{figure}[H]
\centering
\includegraphics[width=0.9\columnwidth]{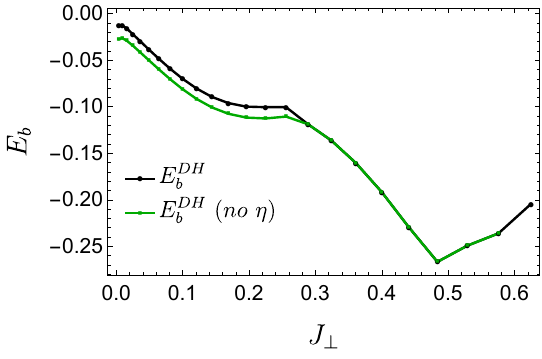}
\includegraphics[width=0.94\columnwidth]{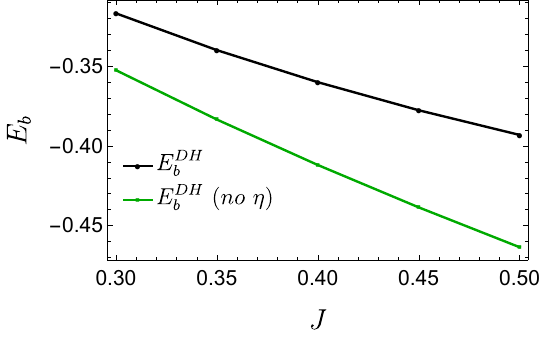}
\caption{Comparison of binding energies for the ladder (left) and 2D geometry (right) in the presence/absence of $\eta$ term in the effective Hamiltonian.}
\label{fig::etaBinding}
\end{figure}

Fig.~\ref{fig::etaBinding} shows that adding the $\eta$-pairing term has a somewhat negative impact on the formation of exciton pairs. It tends to make the binding energy less negative thus implying less strongly bound pairs. This highlights that the role of spin and $\eta$ fluctuations is not equal and that the binding energy does not simply double as we include $\eta$ terms. Qualitatively speaking, results with and without $\eta$ terms are comparable.

\end{document}